\documentclass[preprint]{aastex}

\newcommand{\etal}{et al.\ }
\newcommand{\kms}{\, {\rm km\, s}^{-1}}
\newcommand{\mpc}{\, {\rm Mpc}}

\newcommand{\hmpc}{\, h^{-1} \mpc}

\newcommand{\lya}{Ly$\alpha$ }
\newcommand{\gmo}{\gamma-1}
\newcommand{\bF}{\bar{F}}

\newcommand{\dg}{\delta_g}
\newcommand{\dm}{\delta_m}
\newcommand{\dF}{\delta_F}
\newcommand{\dtm}{\tilde{\delta}_m}
\newcommand{\dtF}{\tilde{\delta}_F}

\slugcomment{Submitted to ApJ}

\shorttitle{Galaxy-Ly$\alpha$ Forest Correlation}

\shortauthors{McDonald \etal}

\begin{document}

\title{Large-scale Correlation of Mass and Galaxies with the
\lya Forest Transmitted Flux}

\author{Patrick McDonald\altaffilmark{1,2},
Jordi Miralda-Escud\'e \altaffilmark{2}, and Renyue Cen\altaffilmark{3}
}
\altaffiltext{1}{Department of Physics, Princeton University,
Princeton, NJ 08544; pmcdonal@feynman.princeton.edu}
\altaffiltext{2}{Department of Astronomy, The Ohio State University, 
Columbus, OH 43210; jordi@astronomy.ohio-state.edu}
\altaffiltext{3} {Princeton University Observatory, 
Princeton University, Princeton, NJ 08544; cen@astro.princeton.edu}

\begin{abstract}

  We present predictions of the correlation between the \lya 
forest absorption in quasar spectra and the mass 
within $\sim 5 \hmpc$ (comoving) of the line of sight, 
using fully hydrodynamic and 
hydro-PM numerical simulations of the cold dark matter model 
supported by present observations. 
The observed correlation based on galaxies
and the \lya forest can be directly compared to our theoretical results,
assuming that galaxies are linearly biased on large scales.
Specifically, we predict the average value of the mass fluctuation,
$<\! \dm \! >$, conditioned to a fixed value of the \lya forest
transmitted flux $\dF$, after they have been smoothed over 
a $10 \hmpc$
cube and line of sight interval, respectively. We find that
$<\! \dm \! >/\sigma_m$ as a function of $\dF/\sigma_F$ has a slope 
of $0.6$ at this
smoothing scale, where $\sigma_m$ and $\sigma_F$ are the rms dispersions
(this slope should decrease with the smoothing scale).
We show that this value is largely insensitive to the cosmological
model and other \lya forest parameters.
Comparison of our predictions to observations should provide a
fundamental test of our ideas on the nature of the \lya forest and
the distribution of galaxies, and can yield a measurement of the bias
factor of any type of galaxies that are observed in the vicinity of \lya
forest lines of sight.

\end{abstract}


\section{INTRODUCTION}

  The prevalent theory to explain the \lya forest 
is that the absorption lines arise from density variations in a
photoionized intergalactic medium that originate in the gravitational
evolution of primordial fluctuations. Both semi-analytic models and
numerical simulations (Bi 1993; Cen \etal 1994; Zhang \etal 1995, 1998;
Hernquist \etal 1996; Miralda-Escud\'e \etal 1996; Bi \& Davidsen 1997)
have shown that the predicted \lya spectra appear remarkably similar
to the observations. The good agreement of the predicted and observed
flux distribution and power spectrum of the \lya forest (Rauch \etal
1997; Croft \etal 1999; McDonald \etal 2000), and the large transverse
size of the absorption systems (Bechtold \etal 1994; Dinshaw \etal
1994, 1997; Petitjean \etal 1998; Monier, Turnshek, \& Hazard 1999;
Dolan et al. 2000; L\'opez, Hagen, \& Reimers 2000) are the basic 
tests that have so far been done and have
supported the theory. In addition, this \lya forest theory is derived
from the general Cold Dark Matter (hereafter, CDM) model with parameters
that are well constrained from several other observations 
(e.g., Primack 2000).

  Another important test of our ideas of the \lya forest can be done
by observing the correlation of the transmitted flux in a spectrum
with galaxies. Some of these observations have already been done for
small scales and high column density absorbers, which have shown the
expected strong correlation between galaxies and gas halos
(e.g., Bergeron \& Boiss\'e 1991; Steidel, Dickinson, \& Persson 1994;
Lanzetta \etal 1995; Chen \etal 2001).
Penton, Stocke, \& Shull (2001) have probed this correlation
at low redshift, and have found that the weak absorption lines are
often found in low-density regions of the galaxy distribution.
Recently, Adelberger \etal (2001) have carried
out the first analysis of this correlation on large scales (several
comoving Mpc) and at high redshift (using galaxies detected with the
Lyman break technique), using the transmitted flux as the
quantity to correlate with the mean number of galaxies in a specified
region.

  Inspired by the observational results of Adelberger \etal (2001), this
paper presents detailed theoretical predictions for statistical
functions similar to the one introduced by those authors. 
Namely, we analyze the
mean value of the mass given an observed value of the transmitted flux,
and the mean value of the transmitted flux for a fixed value of the
mass, after both quantities have been smoothed over a certain region.
The main difference between the functions we analyze and the function
shown by Adelberger \etal (2001) is that the observed objects are of
course galaxies, and our simulations predict only the distribution of
the mass. However, the predictions of the simulations can still be
compared to the galaxy observations to determine the relation between
galaxies and mass, which is especially straightforward if linear bias is
a sufficient description of the distribution of galaxies relative to the
mass on the large scales being probed. We examine the dependence of
these statistical functions on the various parameters affecting the \lya
forest in \S 3.

\section{METHOD}

  We use several HPM (hydro-particle-mesh; see Gnedin \& Hui 1998)
simulations
to model the \lya forest, after testing that the HPM approximation is
sufficient for our purposes by comparing
to a fully hydrodynamic simulation.  The standard HPM simulations
that will be used most often in this paper have box size $40 \hmpc$,
with $512^3$ particles.
The cosmological model is CDM with cosmological constant, in a flat
universe with present matter density $\Omega_m=0.4$, power spectrum
index $n=0.95$, and
amplitude given by $\sigma_8=0.75$.  All the results in this paper
will be shown at $z=3$, the typical redshift at which the observations
have so far been done. The model we use is consistent with the
observed \lya forest power spectrum: at the typical \lya forest scale
of $k=0.008~(\kms)^{-1}$, it has
$\Delta^2_\rho(k)=0.29$ at $z=3$ [where $\Delta^2_\rho(k)$ is the 
contribution per unit $\ln k$ to the 
variance of the linear theory mass density fluctuations], while the
observed value is $\sim 0.26$ (McDonald et al. 2000; Croft et al. 2002).
The redshift interval corresponding to $40\hmpc$ in this model 
near $z=3$ is $\Delta z = 0.068$.
The density-temperature
relation assumed for the gas is $T(\Delta)=T_{1.4}
(\Delta/1.4)^{\gamma-1}$, where $\Delta$ is the gas density divided by
the mean gas density. We use the parameters $T_{1.4}=17000$ K and
$\gamma-1=0.3$ in our standard simulation (the reason we specify the
value of the temperature $T_{1.4}$ at $\Delta=1.4$ is because the error
in the temperature was determined to be smallest at this density in
McDonald \etal 2002).
Several other similar simulations are used where we vary each one of
the relevant parameters (see \S 3), and we vary also the box size and
resolution to verify the numerical convergence of the results.
Unless otherwise indicated, all the results we show for a given set 
of parameters are averages of four simulations with different random 
initial perturbations.

\subsection{Definitions}

  We define the quantity $\dF \equiv 1 - F/\bar F$, where $F$ is
the fraction of transmitted flux in the \lya forest after the spectrum
has been smoothed along the line of sight with a filter, and $\bar F$ 
is the mean transmitted flux. We use a Gaussian smoothing filter in
this paper, which is $W(kR)= \exp[-(kR)^2/2]$ in Fourier space. We also
define $\dm$ as the redshift-space mass density perturbation,
again smoothed over a specified region. We choose to smooth the mass
fluctuation over cubes of $10 \hmpc$ (comoving)
in redshift space, centered
on the \lya forest line of sight, which is
approximately the smoothing that is used in 
Adelberger \etal (2001).
This is convenient
to do in observations of galaxies, since galaxies are usually
searched for in a square field of view, and they can simply be divided
into redshift bins. 
Note that the size of our cubes in velocity units,
$1280 \kms$, is more fundamental than the size in $\hmpc$, because
the temperature and observed flux power spectrum are both measured in
$\kms$.
 
  In general, the information that can be recovered from observations is
the full joint probability distribution $P(\dF, \dg)$, where $\dg$ is the
fluctuation in the number of galaxies. In this paper we present results
from HPM numerical simulations for $<\! \dm | \dF \! >$, which is the
mean value of the mass fluctuation subject to the condition of a fixed
value in the transmitted flux fluctuation, and for the analogous
quantity $<\! \dF | \dm \! >$. The observed galaxies may not follow
exactly the same fluctuations as the mass, even when smoothed over a
scale of $10 \hmpc$. However, if galaxies are linearly biased at this
large scale, then $\delta_g=b\, \dm$, where $\delta_g$ is the
fluctuation in the galaxies and $b$ is the bias factor. If we now define
the quantity $\dtm=\dm/\sigma_m$, and $\tilde\delta_g=\delta_g/ \sigma_g$,
where $\sigma_m$ and $\sigma_g$ are the rms fluctuation of $\delta_m$
and $\delta_g$, then $\dtm=\tilde\delta_g$ and the correlation with
$\dF$ is precisely predicted. For convenience, we also define the
quantity $\dtF=\dF/\sigma_F$, and we will present results for $<\! \dtm
| \dtF \! >$ and $<\! \dtF | \dtm \! >$. 

\subsection{Smoothing}

  To choose a value of the smoothing radius $R$ for the smoothing
filter of the \lya forest transmitted flux, we compute first the
correlation coefficient $ <\! \dF\dm \! > / \sigma_F\sigma_m$ in
our standard simulation, and show it as a function of $R$ in Figure 1.
The correlation is maximum at $R=3.5 \hmpc$. The dependence of the
correlation on $R$ is not surprising: if $R$ is too small, the value
of $\dF$ is altered by small-scale fluctuations that are not related
to the value of $\dm$ and therefore act as noise, and if $R$ is too
big, then the fluctuations affecting $\dm$ are erased by smoothing in
the value of $\dF$.
\begin{figure}
\plotone{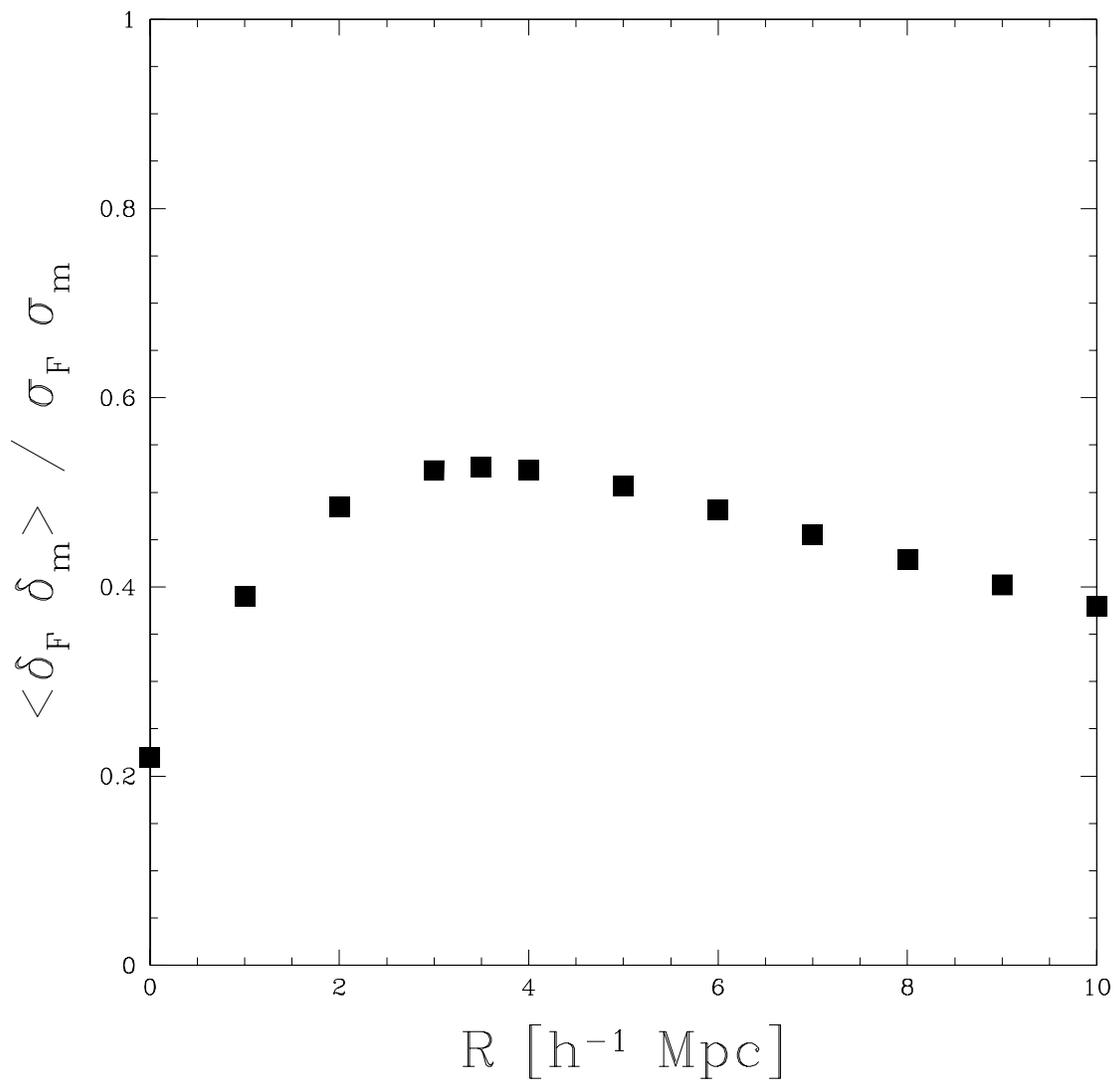}
\caption{Correlation between mass and transmitted flux as a function of
the smoothing applied on the spectrum, with the Gaussian filter,
$W(k R) = \exp[-(k R)/2]$.  The squares show the correlation, 
$<\! \delta_F ~\delta_m\! > /(\sigma_F~\sigma_m$).}
\label{optR}
\end{figure}

   A Gaussian filter with $R=3.5 \hmpc$ will be used for the 
smoothing of the \lya spectrum throughout the rest of this paper.
However, before proceeding, we examine how much our results
differ if we use instead a top-hat smoothing for the \lya spectrum,
or if we vary the size of the cube over which the mass is smoothed.

 Throughout the paper, our results will be presented as a set of
four figures showing the four functions
$<\! \dm | \dF \! >$, $<\! \dF | \dm \! >$,
$<\! \dtm | \dtF \! >$, and $<\! \dtF | \dtm \! >$.
In all these figures, the function $<\! \dm | \dF \! >$
has been obtained by creating spectra along each row of cells in the
simulation, smoothing the spectra along the line of sight, 
selecting all the
pixels in the spectra where the value of $\dF$ is inside a given bin,
and computing the average $\dm$ for these pixels, where $\dm$ is
computed in redshift-space. We use bins of width
$\Delta \dF=0.05$. Similarly, $<\dF |\dm>$ is calculated by averaging
the values of $\dF$ for pixels in bins with a given $\dm$ (again using
$\Delta \dm=0.05$ for the bin width).
\begin{figure}
\plotone{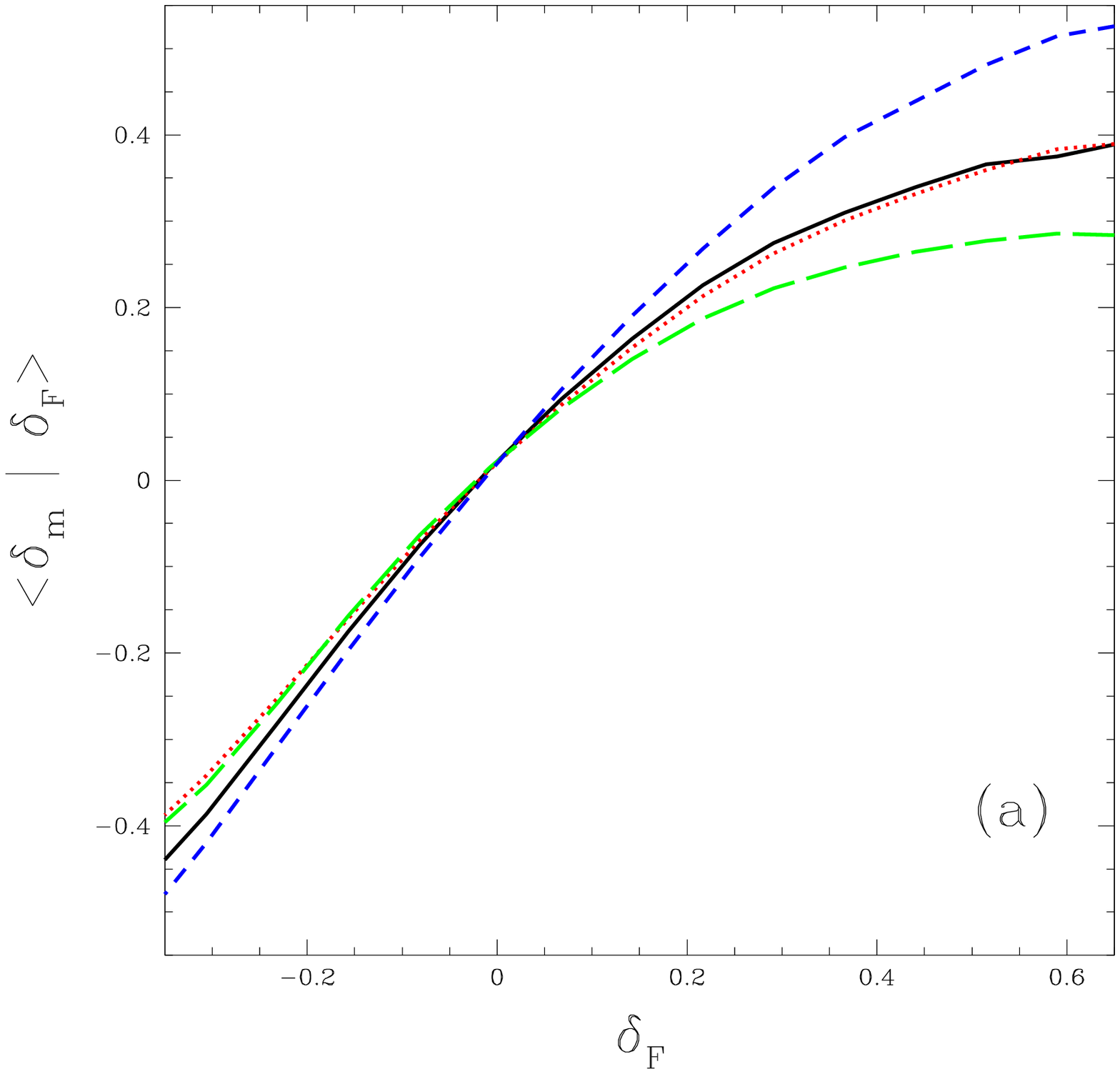}
\caption{ Comparison between
our standard $3.5\hmpc$ Gaussian
smoothing of the spectrum with $10\hmpc$ cube smoothing of the
mass ({\it solid, black line}) 
and $5\hmpc$ top-hat smoothing of the spectrum
({\it dotted, red line}), $12\hmpc$ cube smoothing of the mass
with $4.2 \hmpc$ Gaussian smoothing of the spectrum
({\it long-dashed, green line}), or $8\hmpc$ cube smoothing of the mass
with $2.7 \hmpc$ Gaussian smoothing of the spectrum
({\it short-dashed, blue line}). 
(a) mean mass as a 
function of transmitted flux, (b) mean flux as a function 
of mass, (c and d) are as (a and b) except that $\dF$ and $\dm$
are divided by their rms fluctuation. The mass is always computed in
redshift space.}
\label{smoothcomp}
\end{figure}
\begin{figure}
\plotone{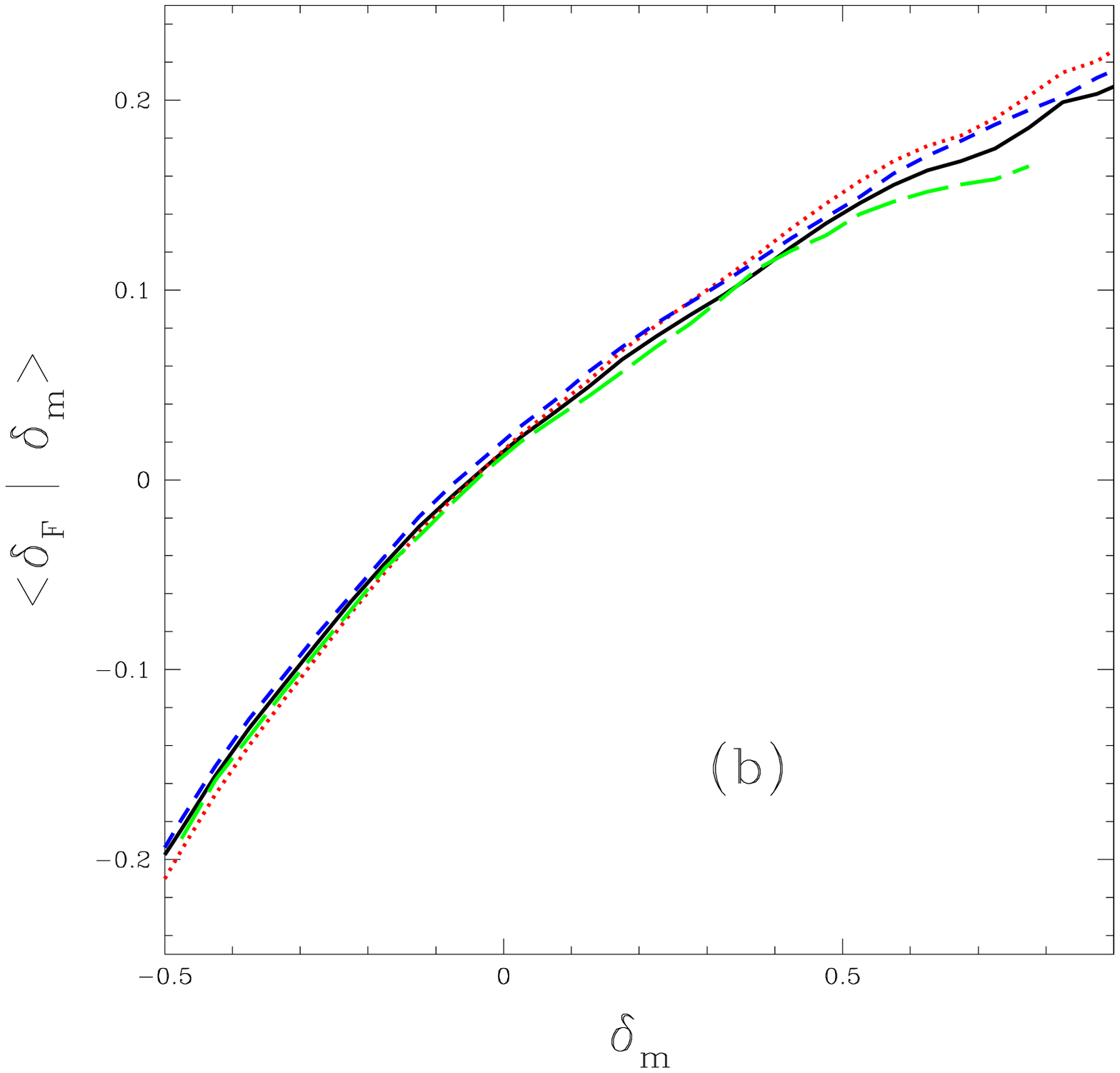}
\end{figure}
\begin{figure}
\plotone{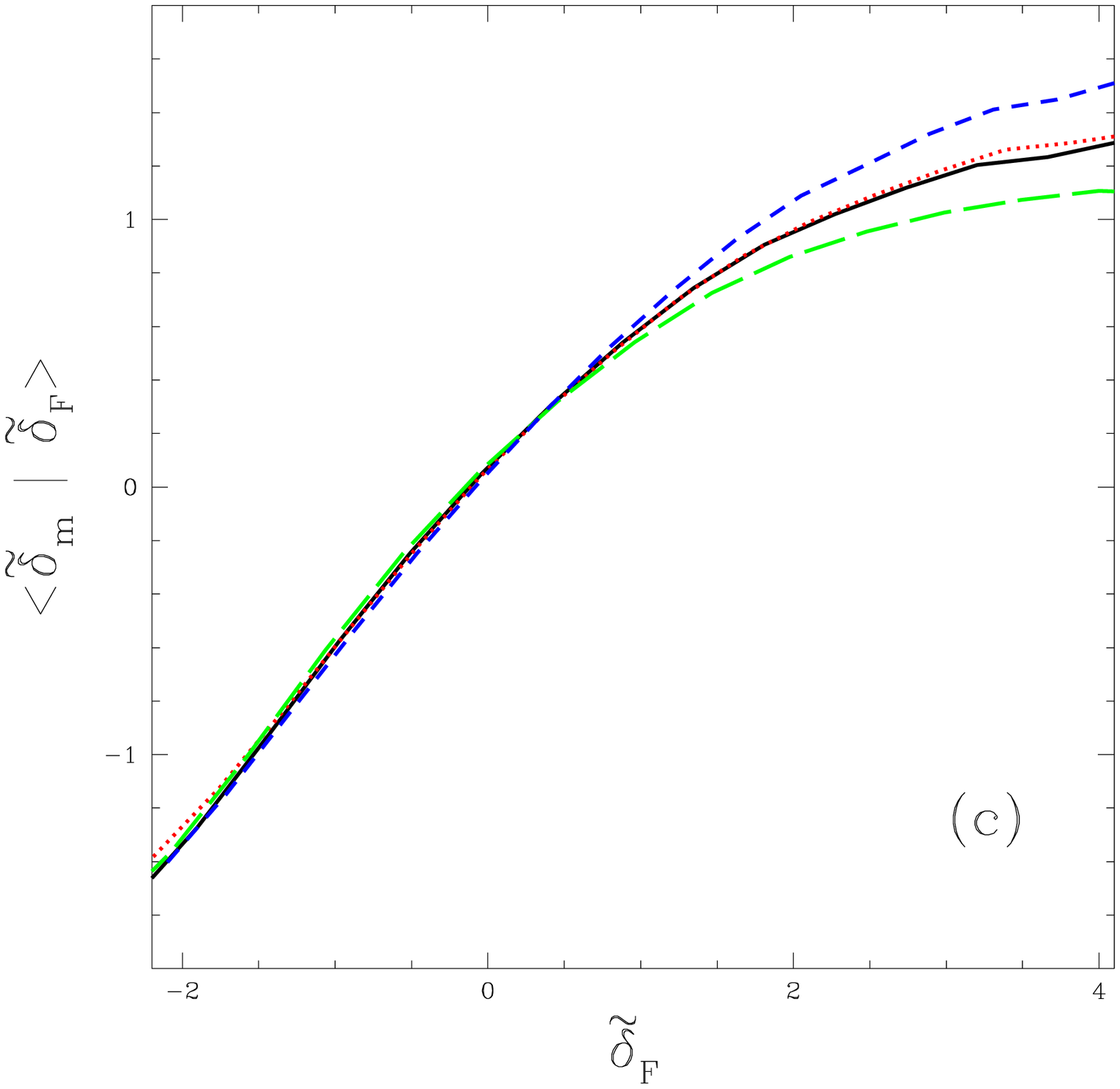}
\end{figure}
\begin{figure}
\plotone{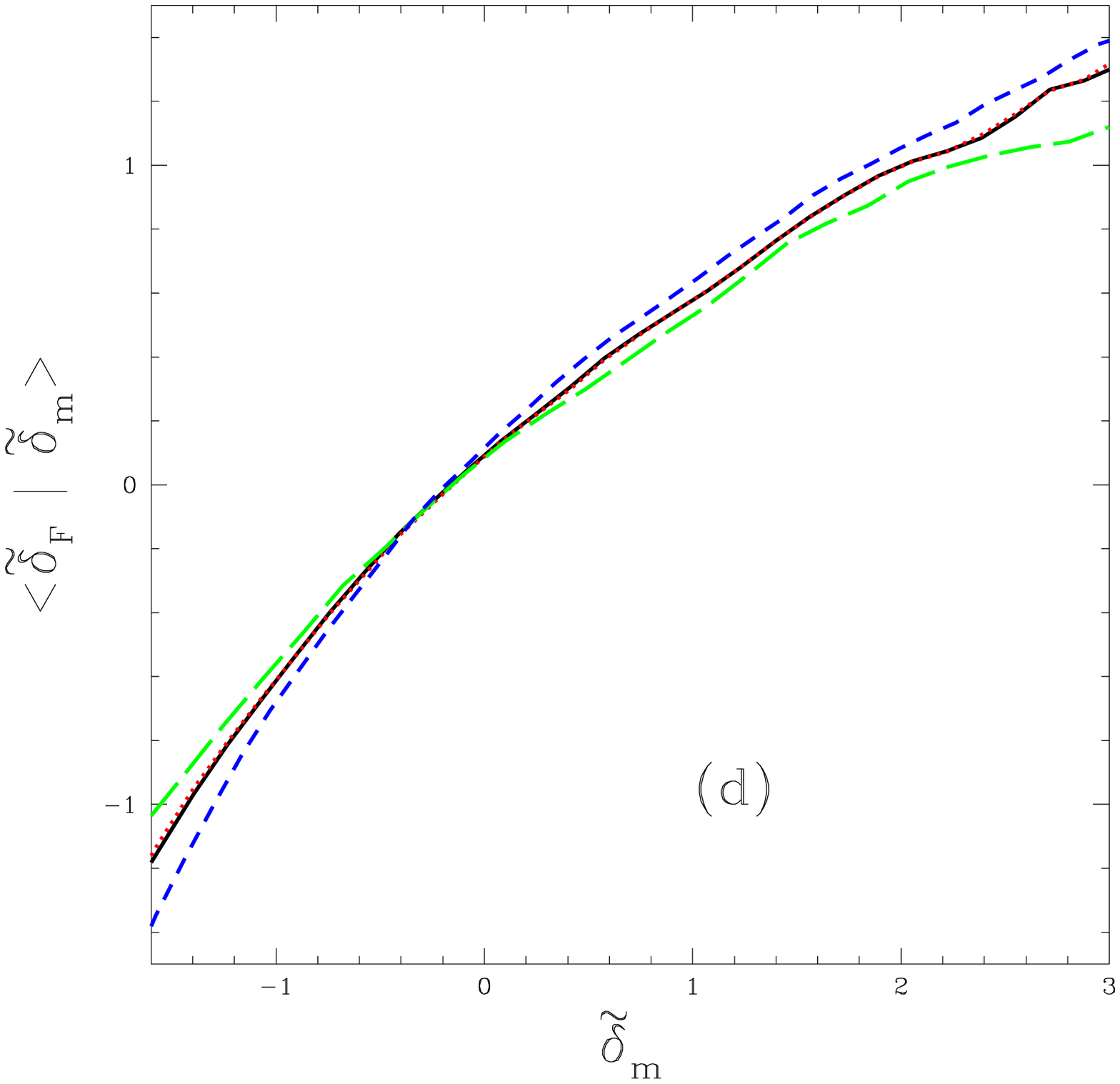}
\end{figure}

 In Figure \ref{smoothcomp}(a-d), we compare the 
results for our standard Gaussian filter
(black, solid line)
to the results obtained if we simply average the \lya forest flux
across the $10\hmpc$ extent of the cube with which the mass is 
smoothed, i.e., use a top-hat filter with $R=5\hmpc$ 
(red, dotted line).  
The difference
is not large, although noticeable, for the $\dm$-$\dF$ relations,
and is practically zero for the $\dtm$-$\dtF$ relations.

   We also show in Figure \ref{smoothcomp} the effect
of changing the size of the cube used to smooth the mass density
field from $10\hmpc$ to $12 \hmpc$ (green, long-dashed line)
or $8\hmpc$ (blue, short-dashed line).  In both of these cases
we also change the smoothing length for the spectrum by the same
factor.
As the size of the smoothing
cube is reduced, the mass fluctuations increase and they are more
strongly correlated with the \lya forest, as expected. 
We see that for $<\! \dm | \dF \! >$
it is important to match the dimensions of the cube in the simulations
and observations to compare the two.
Note, however, that the transverse angular size of the cube depends 
on the cosmological geometry.  The other plots show less sensitivity
to the cube size.

To promote comparison between our results and observations or other
simulations,
in Table \ref{restab} we give $<\! \dtm | \dtF \! >$ 
and $<\! \dtF | \dtm \! >$ for the three cube sizes, along
with $\sigma_F$ and $\sigma_m$ needed to convert from $\tilde{\delta}$
to $\delta$.  The rows in the Table labeled 2s, 2ld, and 2sd 
correspond to cube size $10\hmpc$, $12\hmpc$, and $8\hmpc$, 
respectively. 

\subsection{Test of the HPM Approximation}

  We resort to approximate HPM simulations (Gnedin \& Hui 1998) because
it is impractical at the present time to perform fully hydrodynamic
simulations of the required size for all of the parameter variations
we would like to explore.  However, we first test the accuracy of the
HPM approximation by comparing to a state of the art hydrodynamic
simulation of a similar cosmological model.
This simulation is Eulerian, with box size $25\hmpc$ 
divided into $768^3$ cells for baryons, with $384^3$ dark matter 
particles (see Cen \etal 2001 for a more complete description). 
We compare to a $25\hmpc$, $512^3$ particle HPM simulation with
identical initial Fourier modes up to the Nyquist frequency of
the mesh (as we show in the next subsection, the resolution
of these simulations is sufficient for convergence of our
chosen statistics, so we do not need to worry about exactly how to 
equate resolution between the two).

\begin{figure}
\plotone{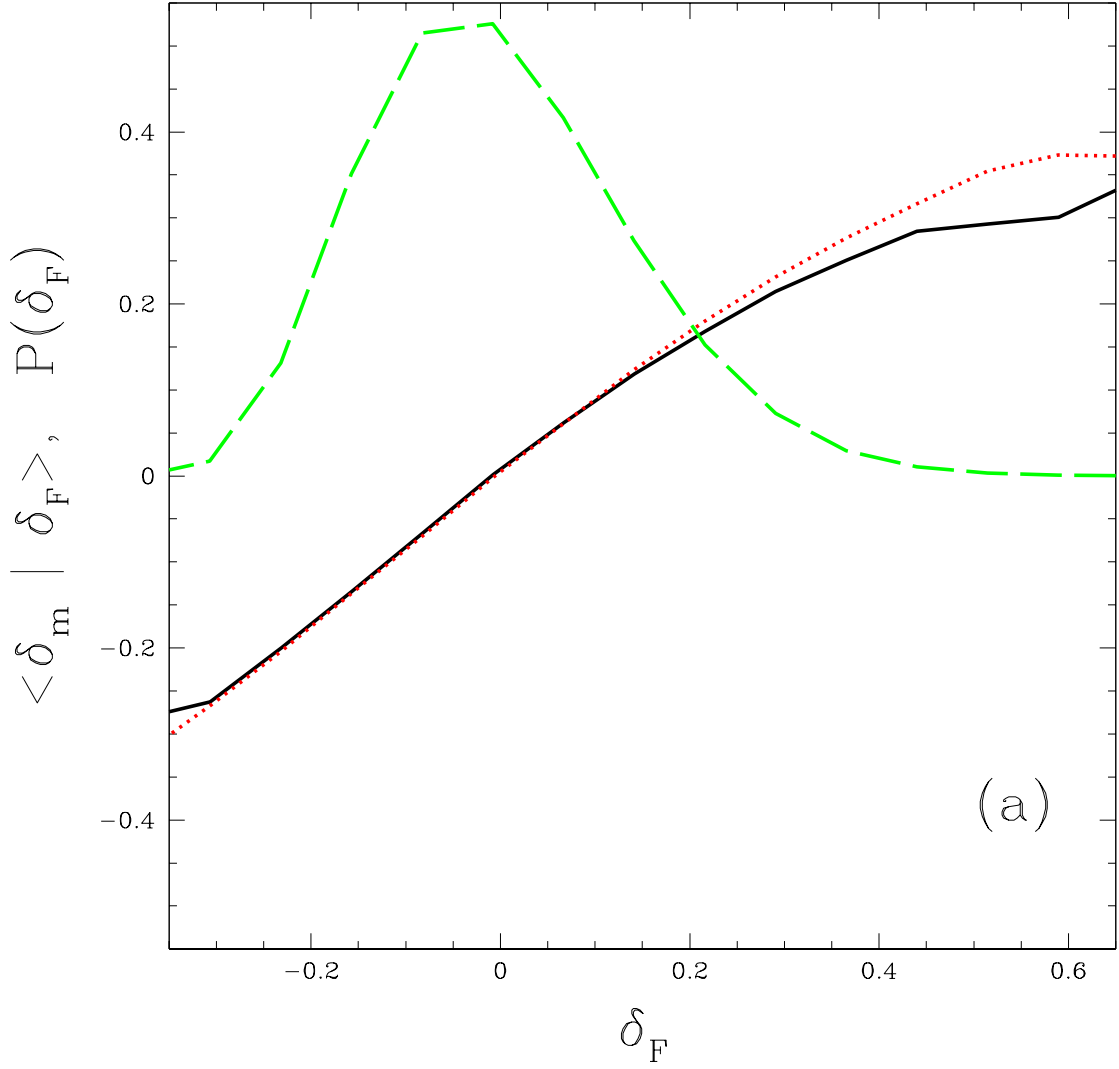}
\caption{Test of the HPM approximation.  
 {\it Black (solid) lines:} HPM results; 
 {\it red (dotted) lines:} fully hydrodynamic results;
 {\it green (dashed) lines:} PDF of $\dF$ (a) and 
 $\dm$ (b).  The normalization of the PDFs is arbitrary.}
\label{hydrocheck}
\end{figure}

  The comparison for our statistics is shown in Figure \ref{hydrocheck},
where the solid lines are the HPM results and the dotted lines are
the fully hydrodynamic results.
\begin{figure}
\plotone{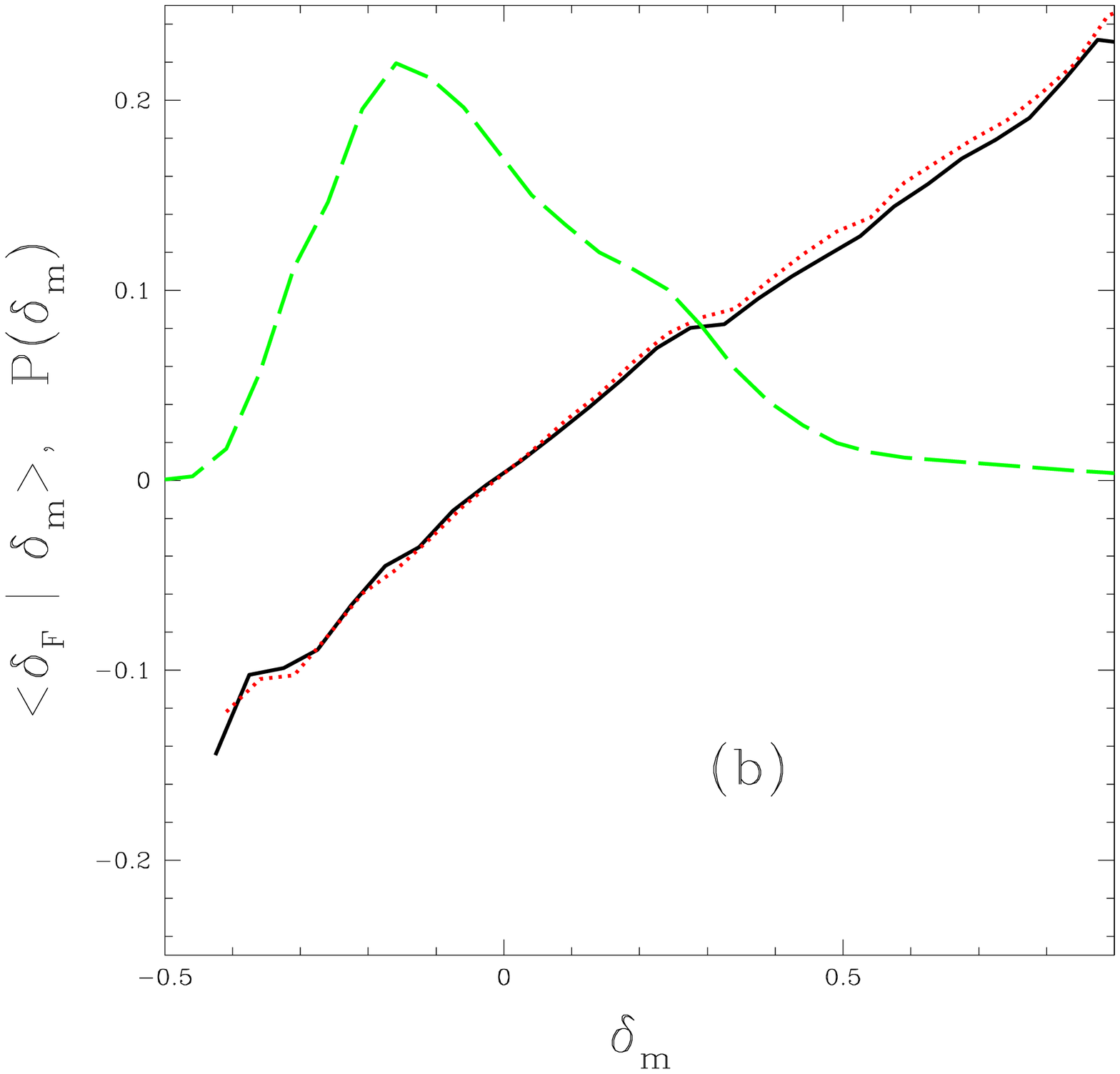}
\end{figure}
\begin{figure}
\plotone{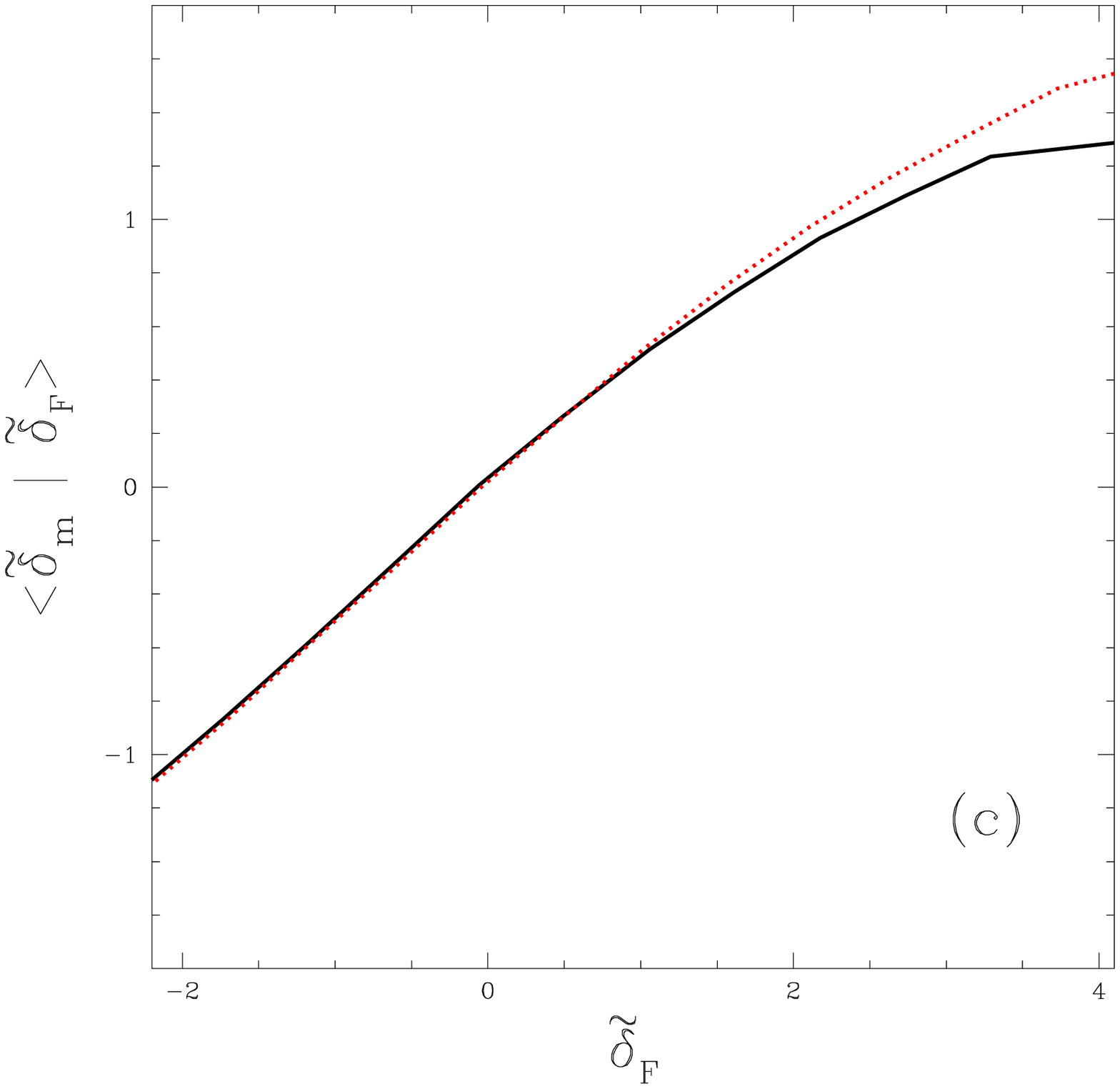}
\end{figure}
\begin{figure}
\plotone{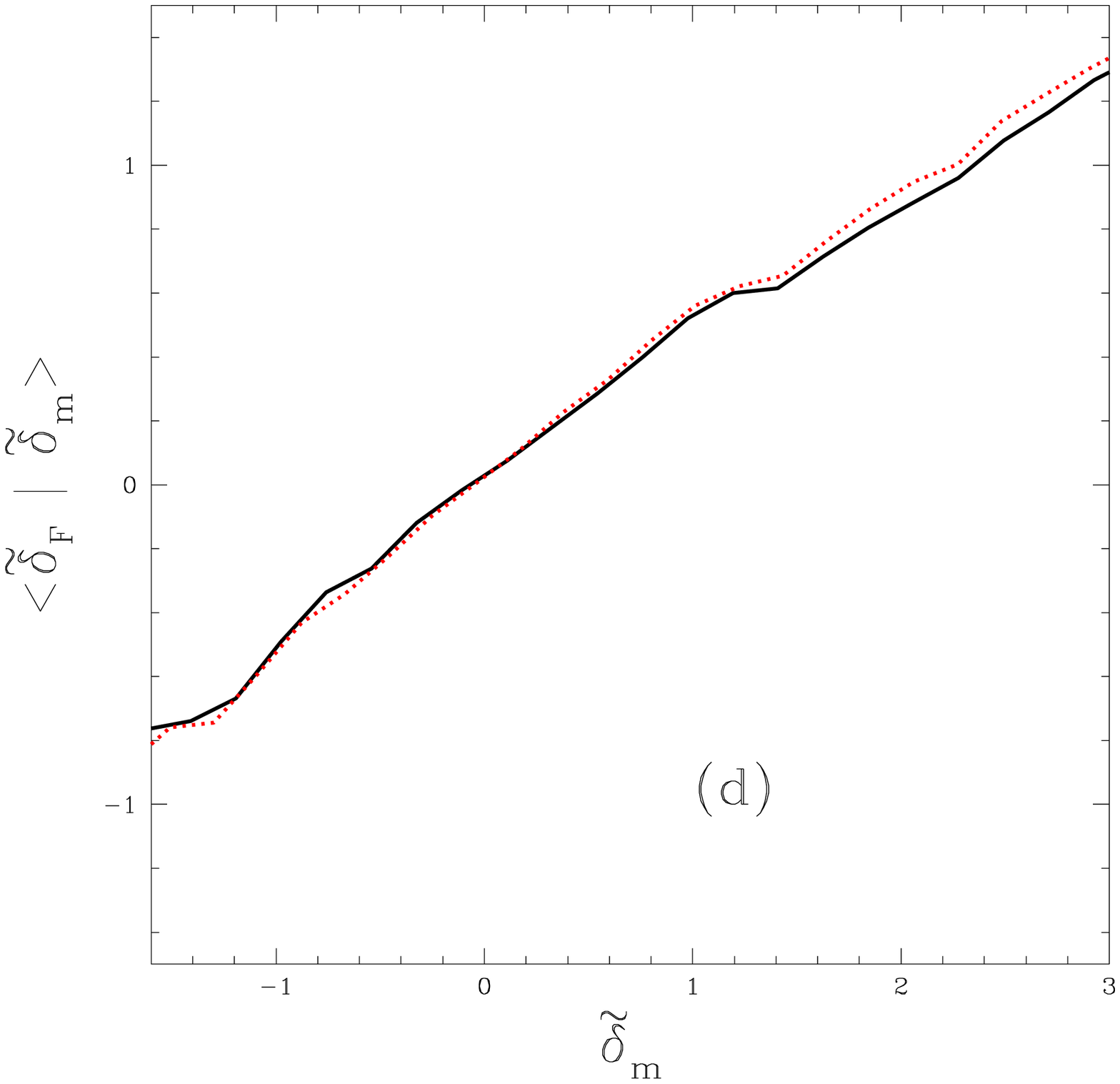}
\end{figure}
The agreement is excellent except in the highest $\dF$ regions
in the $<\! \dm | \dF \! >$ and $<\! \dtm | \dtF \! >$
comparisons (a and c).  
For reference, in Figure \ref{hydrocheck}(a and b) we plot the
PDFs of $\dF$ and $\dm$, respectively (defined as the relative
volume-weighted probability of finding $\delta$ within each of 
our usual bins). 
We see that the regions where the HPM approximation
breaks down are extremely rare (at low $\dm$ in Figure 2b the lines
terminate at the lowest density found in the simulations).

\subsection{Convergence of the Results with Resolution and Box Size}

  We start by testing the sensitivity of our numerical results to the
resolution. For this purpose, we use boxes with size $20\hmpc$
in addition to our standard value of $40\hmpc$.

\begin{figure}
\plotone{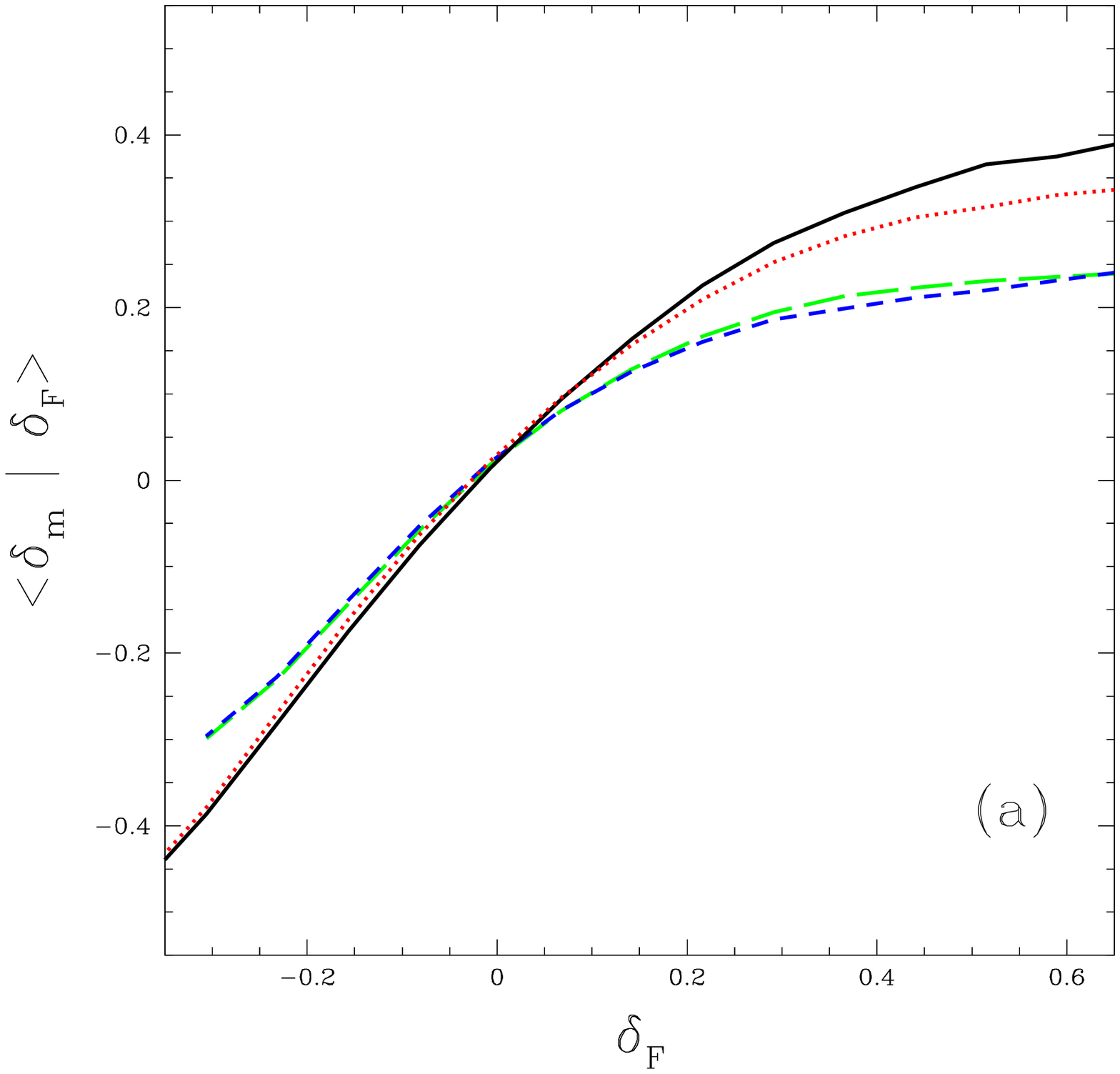}
\caption{Resolution tests.  
{\it Green (long-dashed) line:} $20 \hmpc$ simulation box with
$512^3$ particles; {\it blue (short-dashed) line:}
$20\hmpc$, $256^3$ particles.
{\it Black (solid) line:} $40 \hmpc$, $512^3$ particles; 
{\it red (dotted) line:} $40\hmpc$,
$256^3$ particles. (a) mean mass as a 
function of transmitted flux, (b) mean flux as a function 
of mass, (c and d) are as (a and b) except that $\dF$ and $\dm$
are divided by their rms fluctuation. The mass is always computed in
redshift space.}
\label{rescheck}
\end{figure}

  The results for the four functions, $<\! \dm | \dF \! >$, 
$<\! \dF | \dm \! >$, $<\! \dtm | \dtF \! >$, and $<\! \dtF | \dtm \! >$, 
are shown in Figures \ref{rescheck}(a,b,c,d) for our standard 
cosmological model,
with box size $20 \hmpc$, and resolution of $512^3$ and $256^3$
particles for the long-dashed and short-dashed lines, respectively.
The solid and dotted lines show results with box size $40 \hmpc$,
again with $512^3$ and $256^3$ particles, respectively. 
\begin{figure}
\plotone{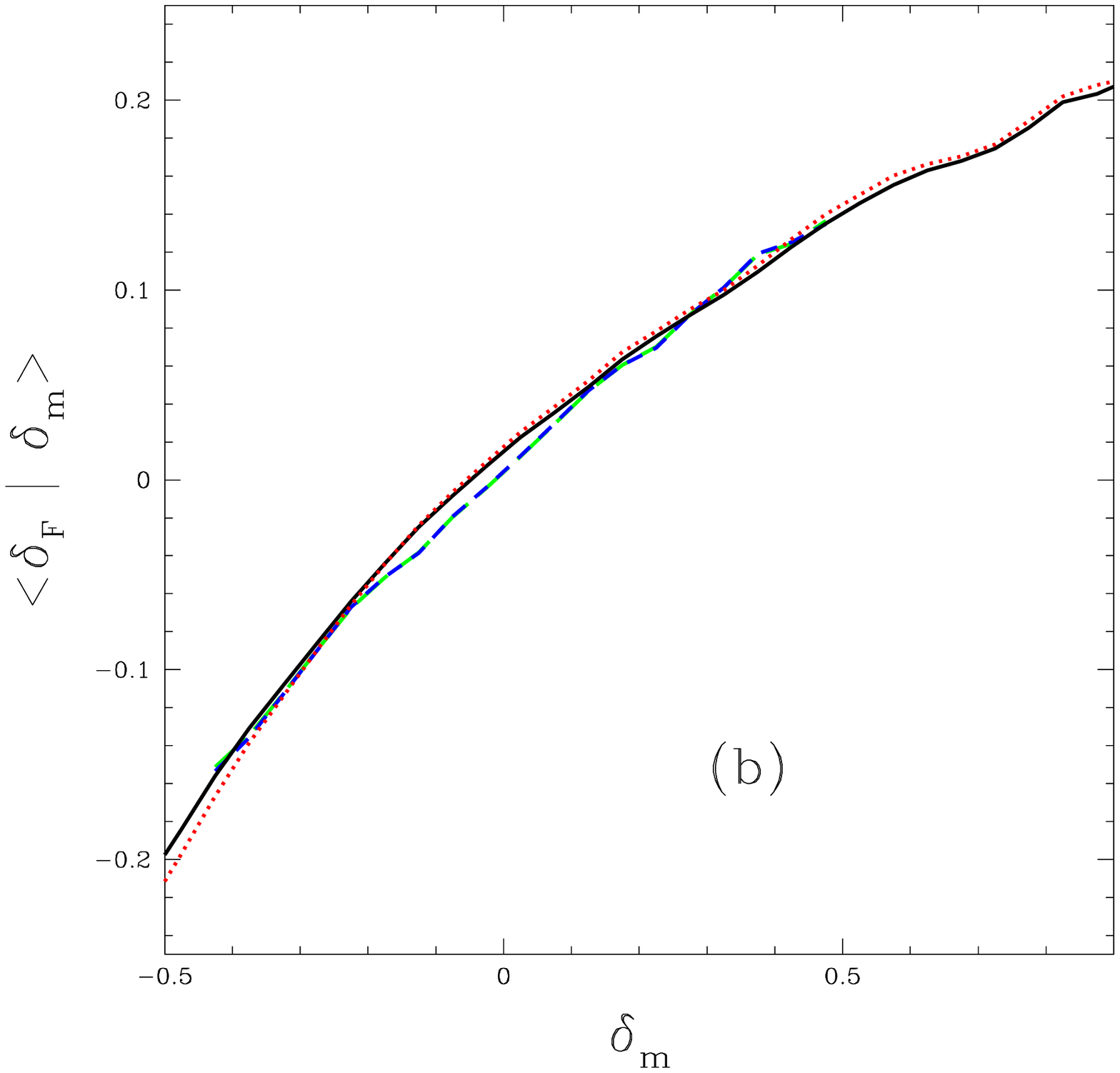}
\end{figure}
\begin{figure}
\plotone{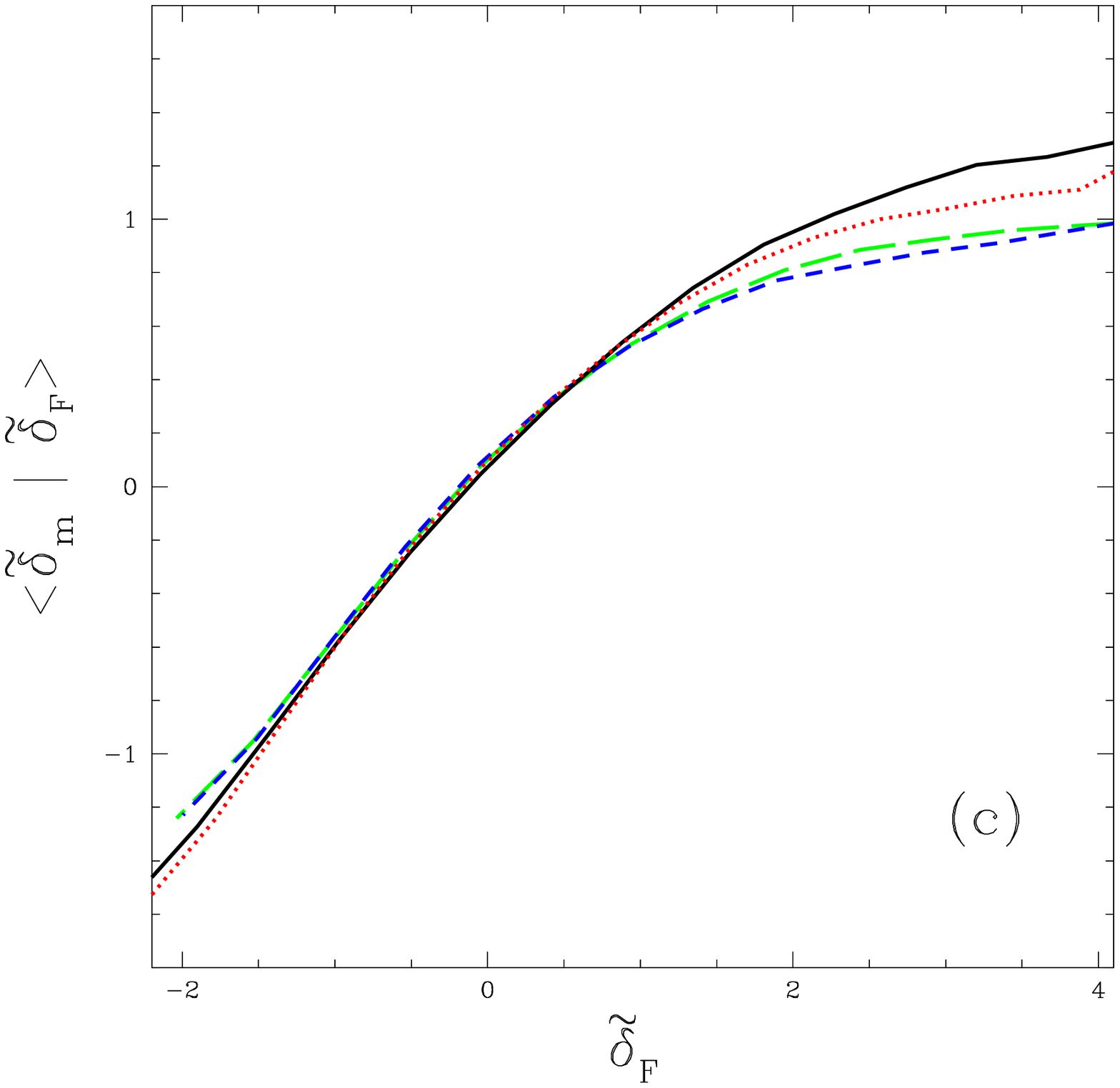}
\end{figure}
\begin{figure}
\plotone{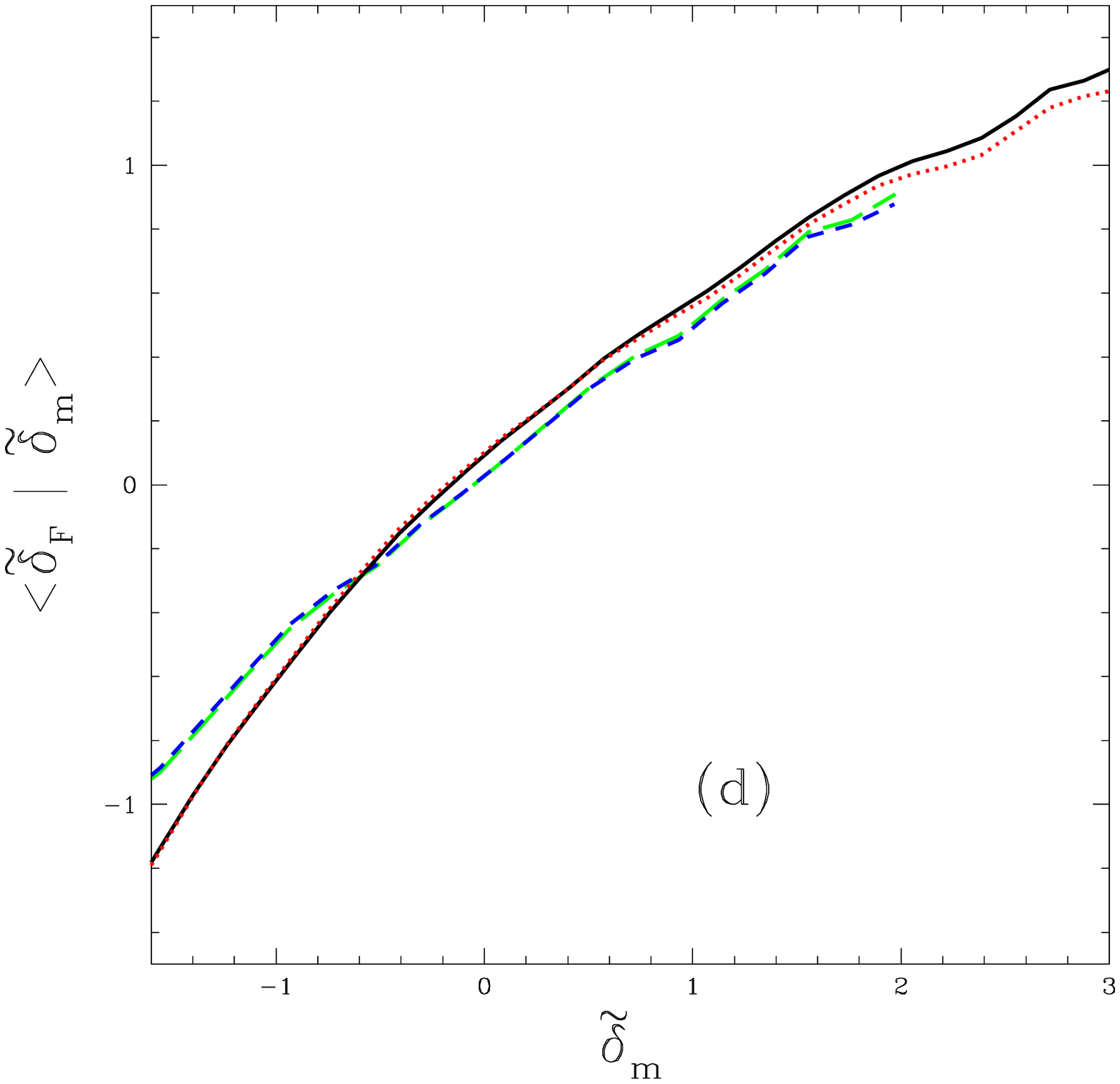}
\end{figure}
The agreement between the $20 \hmpc$ simulations differing only in the
resolution is excellent, verifying that $40 \hmpc$ simulations with 
$512^3$ particles should be well resolved.  The agreement between 
$40\hmpc$ simulations with different resolution is still quite good
except at high $\dF$, indicating that $80 \hmpc$, $512^3$ particle
simulations could still be useful.

\begin{figure}
\plotone{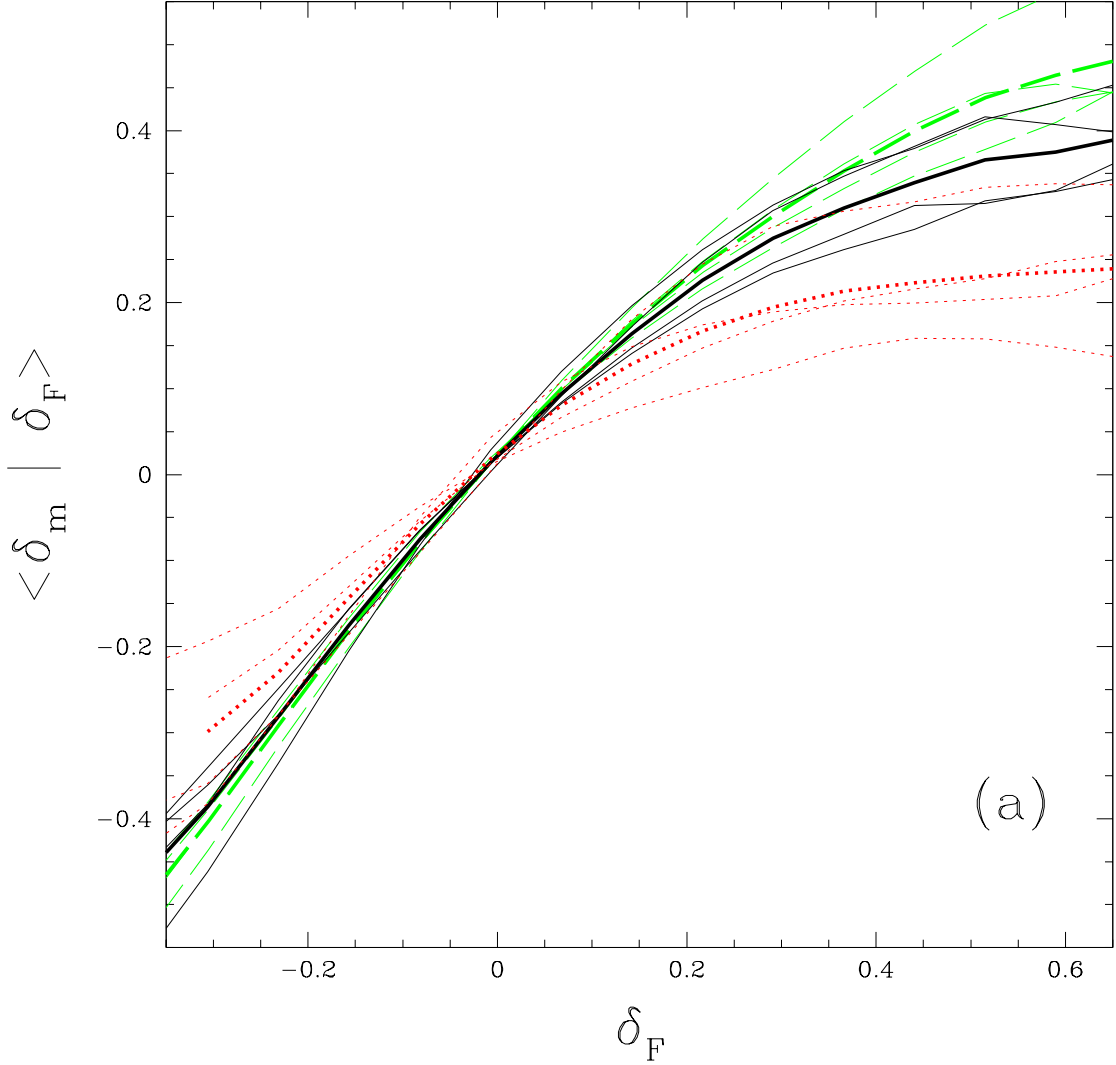}
\caption{Box size and statistical error test.  The thin curves
are from four independent simulations, the thick curves are
the average of the four. {\it Black (solid) lines:} $40 \hmpc$
simulations; {\it red (dotted) lines:} $20 \hmpc$ simulations;
{\it green (dashed) lines:} $80 \hmpc$ simulations.  
All with $512^3$ particles (the $80 \hmpc$ simulations are 
slightly under-resolved).}
\label{boxcheck}
\end{figure}

Figure \ref{rescheck} shows a disturbingly large difference between the 
results for $20\hmpc$ and $40\hmpc$ boxes; however, 
assessing the effect of the box size is more difficult than for the
resolution, because we cannot use the same set of initial mode
amplitudes at each wavenumber, so the statistical fluctuations in
the simulations introduce significant random differences in the result.
Figures \ref{boxcheck}(a-d) compare results for $40 \hmpc$ boxes
({\it black, solid lines}), $20 \hmpc$ boxes ({\it red, dotted
lines}), and $80 \hmpc$ boxes ({\it green, dashed
lines}), all with $512^3$ particles. 
\begin{figure}
\plotone{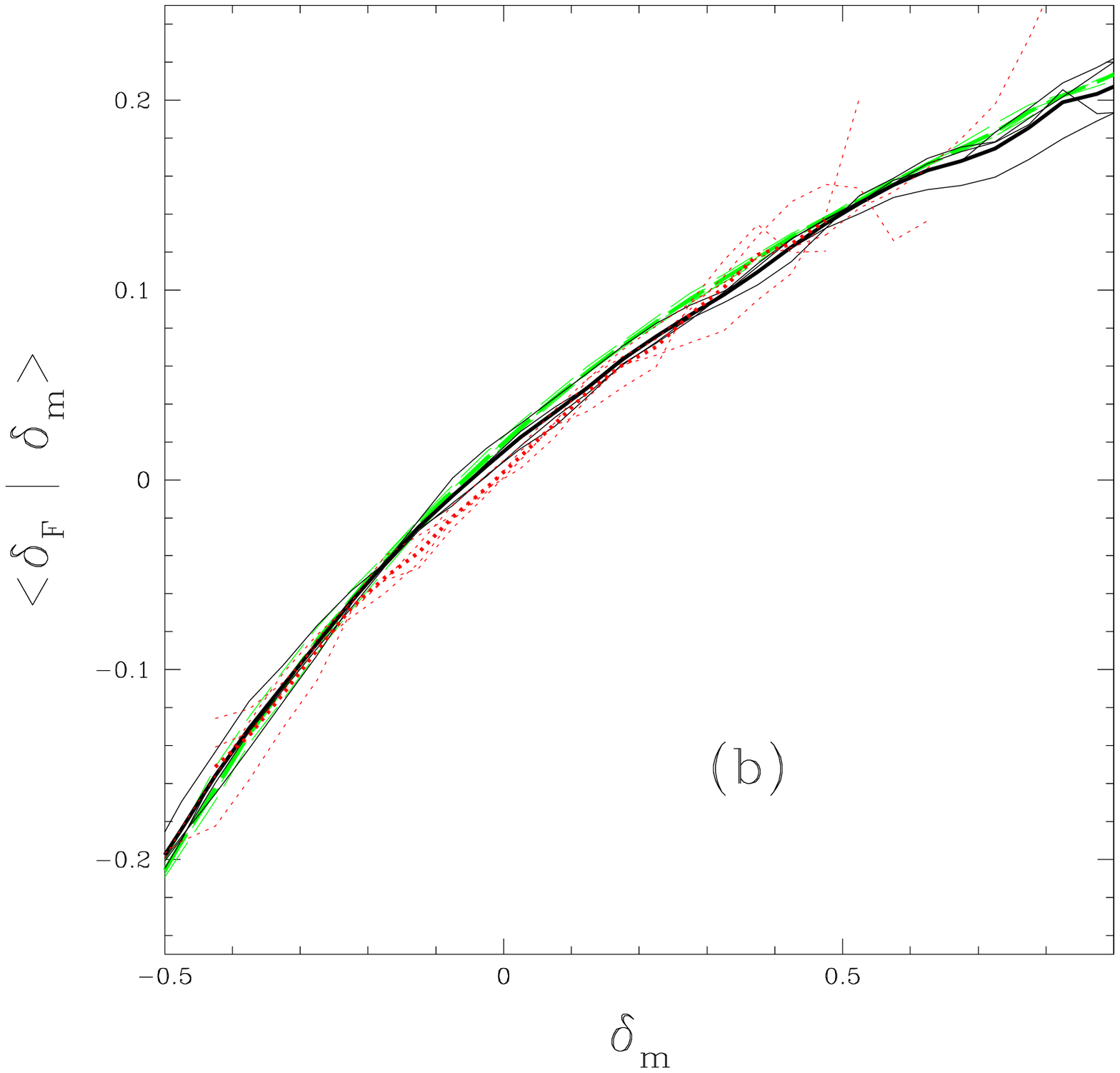}
\end{figure}
\begin{figure}
\plotone{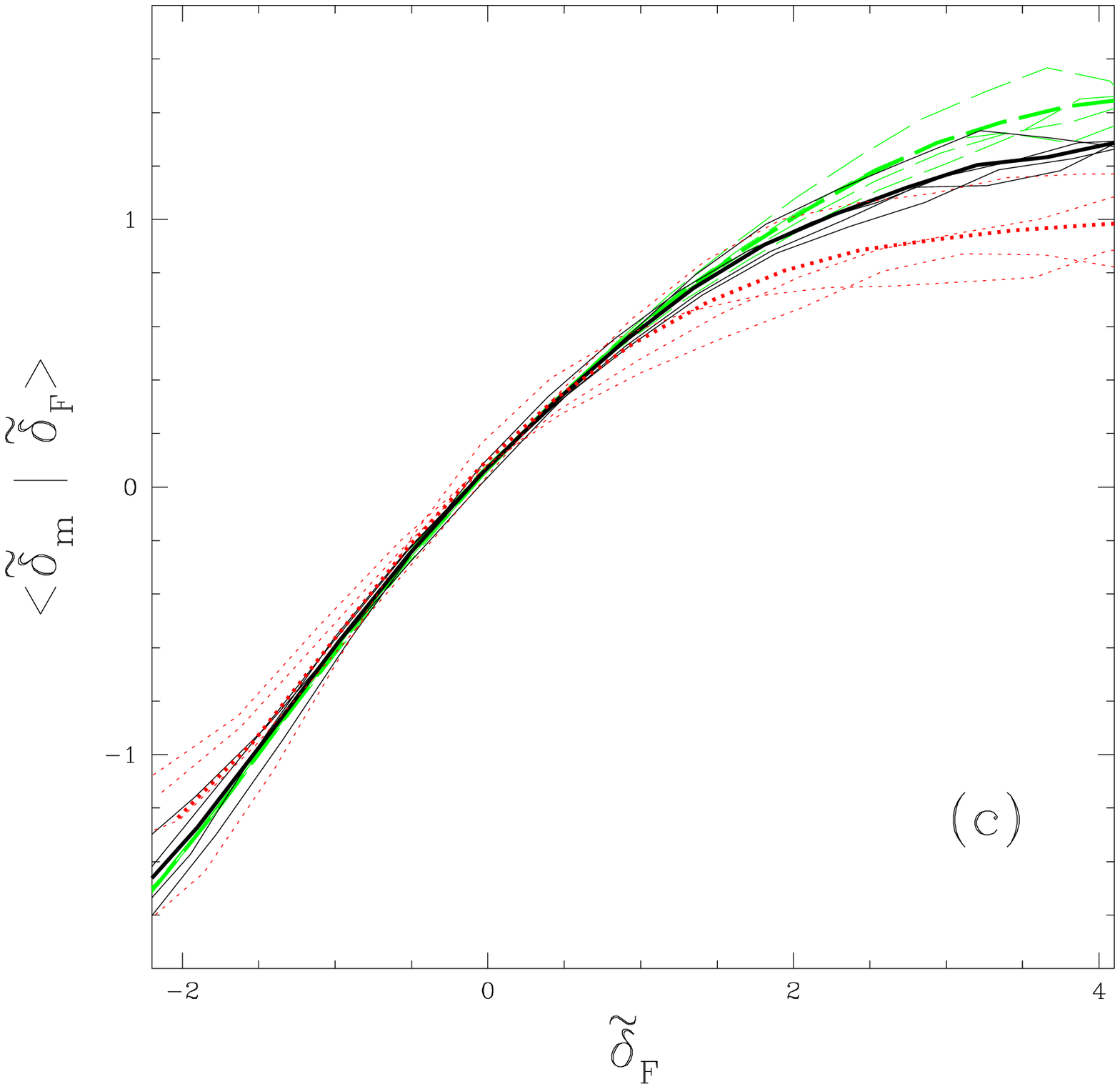}
\end{figure}
\begin{figure}
\plotone{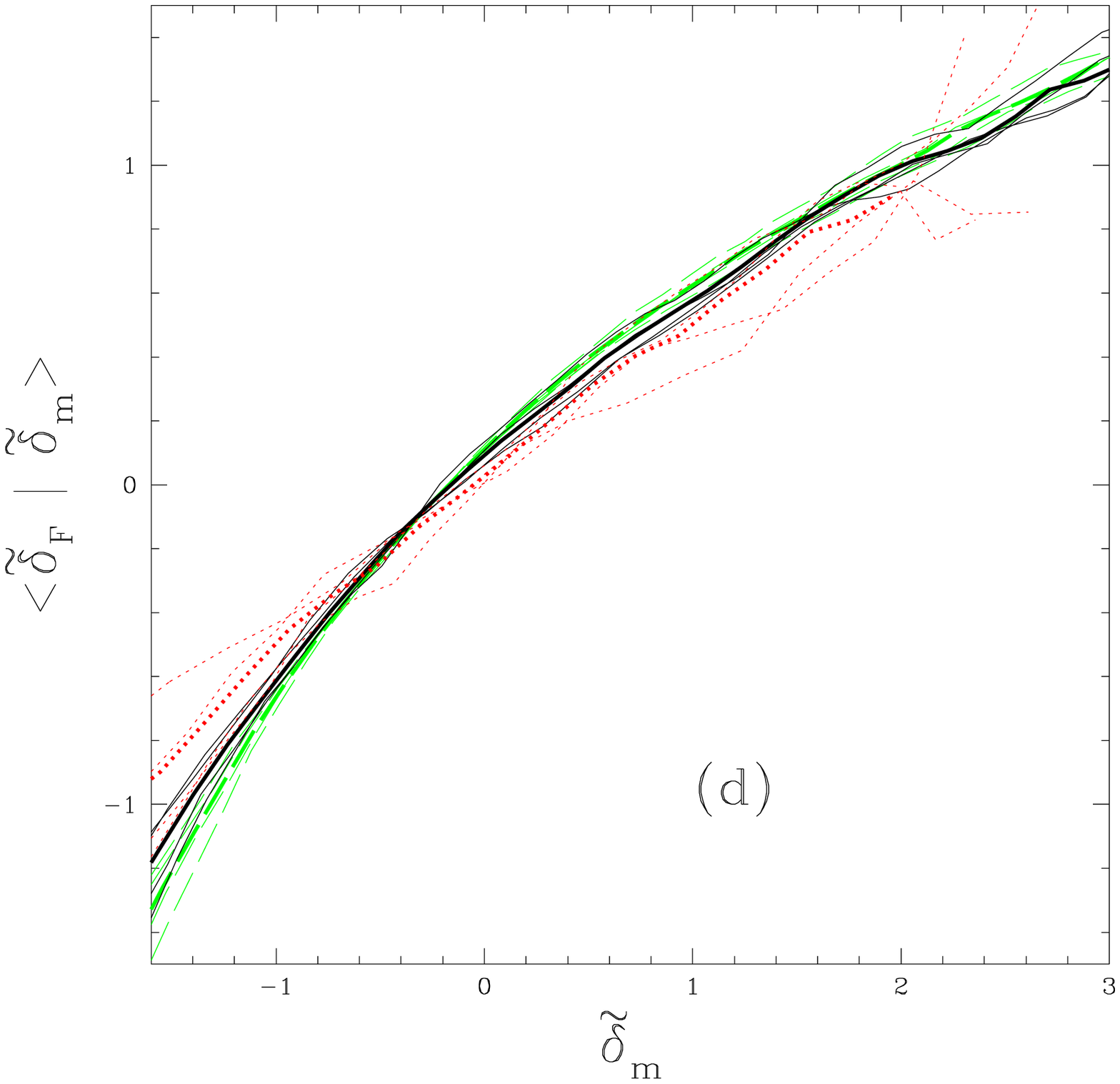}
\end{figure}
The thick lines show the average of four simulations, and the
thin lines show the results from each separate simulation to 
demonstrate the statistical error in the curves.

Figure \ref{boxcheck}(a) shows good agreement between the
40 and $80 \hmpc$ boxes for $<\! \dm | \dF \! >$,
except at large $\dF$, where the trend toward increasing 
mean mass with increasing box size from 20 to 40 to 
$80 \hmpc$ is probably not a result of statistical fluctuations.
In fact, the true difference between the
40 and $80 \hmpc$ boxes is even larger than what is shown because
the decreased resolution in the $80 \hmpc$ box has the effect
of suppressing $<\! \dm | \dF \! >$ at large $\dF$ (see Figure
\ref{rescheck}a).  It is clear that much of the flattening of 
$<\! \dm | \dF \! >$ that we see at large $\dF$ is 
an effect of the finite box size.  Fortunately, the other 
statistics shown in Figure \ref{boxcheck}(b-d) generally 
exhibit better convergence than $<\! \dm | \dF \! >$.

  The results in Figures \ref{boxcheck}(a-d) should also serve as a 
cautionary reminder
that statistical fluctuations from one simulation to another can be
large, and must be taken into account when comparing to observations
or to other simulations, especially single, relatively
small simulations. 
Surprisingly, the scatter
between the separate simulations in the high $\dF$ corner
of the plot is not reduced by increasing the box size from
40 to $80 \hmpc$.  This may be an indication that the statistical
fluctuations in the result are still dominated by the longest
wavelength modes in the box.  

  For reference, for our standard model, in a $40 \hmpc$ box with
our standard smoothing, the rms fluctuations in $\dF$
and $\dm$ are $\sigma_F=0.16$ and $\sigma_m=0.30$.

\section{PARAMETER DEPENDENCE OF THE LY$\alpha$ FOREST - MASS CORRELATION}

  We now examine the dependence of the four functions
$<\! \dm | \dF \! >$, $<\! \dF | \dm \! >$, $<\! \dtm | \dtF \! >$,
and $<\! \dtF | \dtm \! >$ on the most important parameters
determining the properties of the \lya forest. These are the mean
transmitted flux, the mean temperature-density relation of the
intergalactic gas (parameterized as $T=T_0 \Delta^{\gamma-1}$), and
the amplitude and power-law slope of the power spectrum. 
These parameters fully determine
all the physical effects that are incorporated in an HPM simulation
and the calculation of the \lya spectra: the underlying dark matter
fluctuations (assumed to be Gaussian initially), the smoothing of the
gas distribution on the Jeans scale, and the neutral fraction of the
gas and thermal broadening.

\subsection{Mean transmitted flux}

  The value of the mean transmitted flux used in our standard model,
$\bF=0.67$, is appropriate for $z= 3$. The variation of the mean
transmitted flux is the most important factor accounting for the
changes in the \lya forest with redshift.

\begin{figure}
\plotone{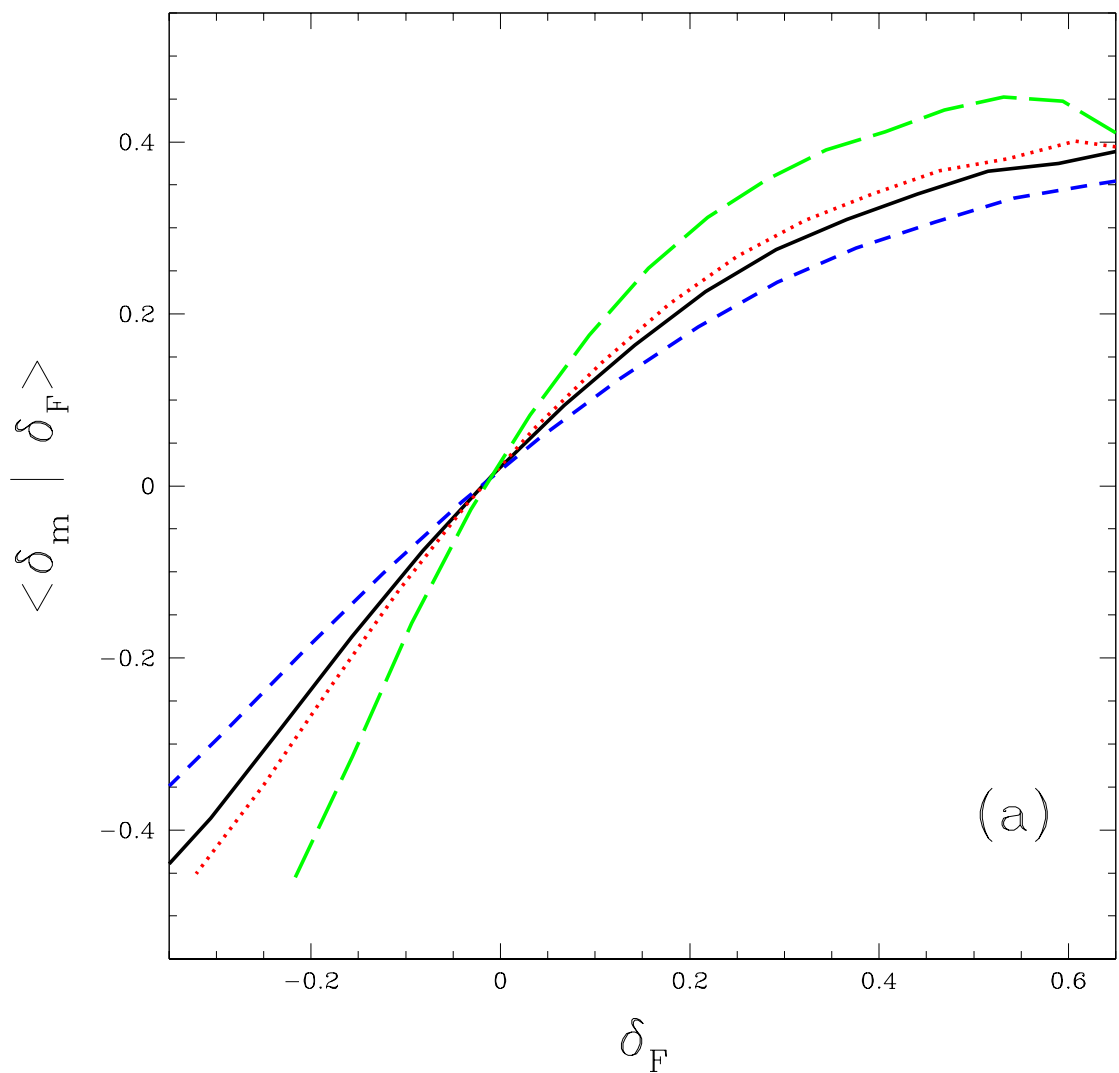}
\caption{Effect of mean transmitted flux fraction.  
Shown are $\bF=0.67$
(black, solid line), $\bF=0.8$, (green, long-dashed line), 
$\bF=0.6$, (blue, short-dashed line), and $\bF=0.7$
(red, dotted line), for the same four functions as previously.}
\label{fluxcheck}
\end{figure}

  Figures \ref{fluxcheck}(a-d) show the four functions defined
previously for $\bF=0.67$ (black, solid line), $\bF=0.8$,
appropriate for $z\simeq 2.5$ (green, long-dashed line), $\bF=0.6$,
appropriate for $z\simeq 3.5$ (blue, short-dashed line), and $\bF=0.7$
(red, dotted line).  
\begin{figure}
\plotone{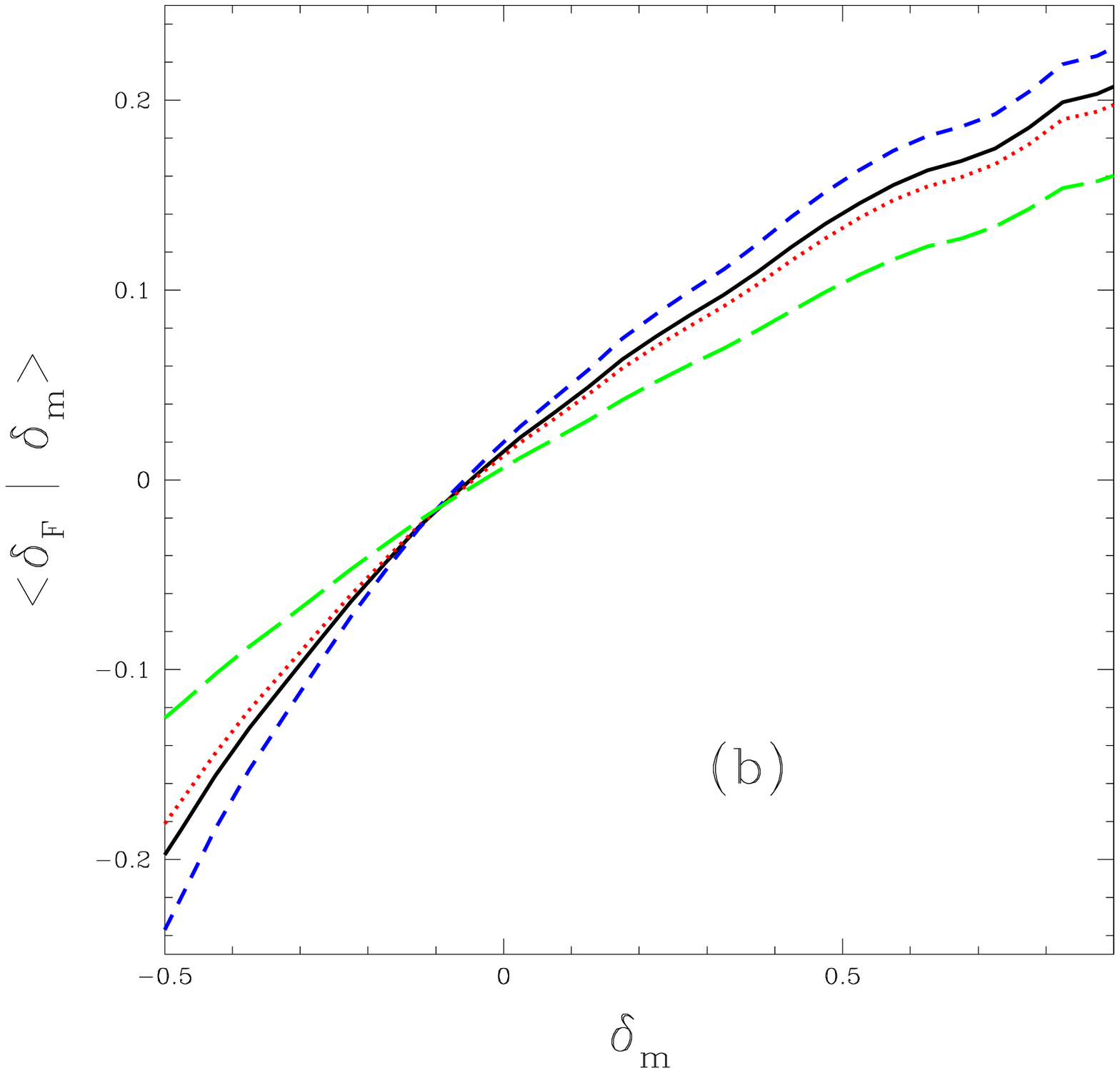}
\end{figure}
\begin{figure}
\plotone{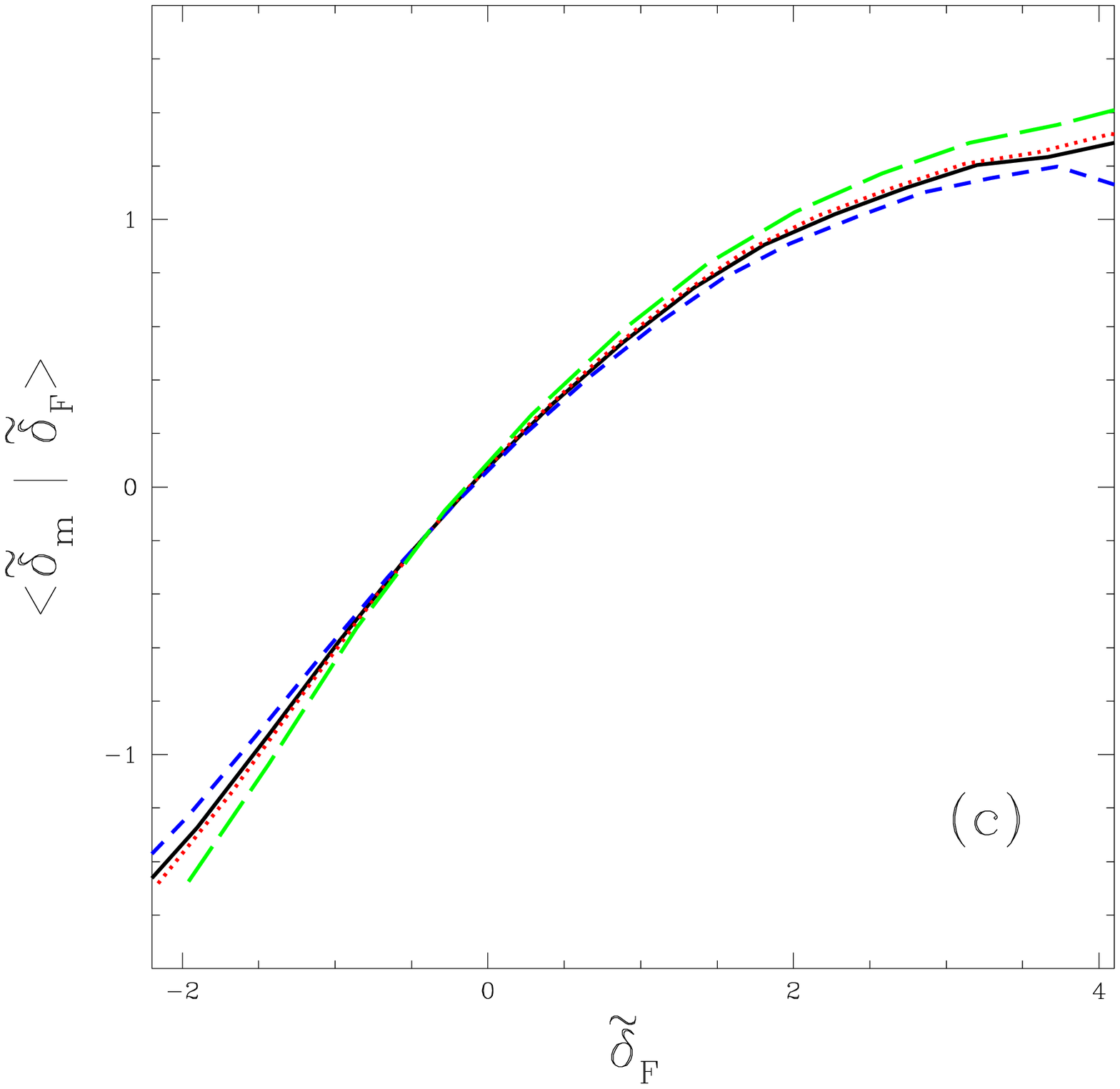}
\end{figure}
\begin{figure}
\plotone{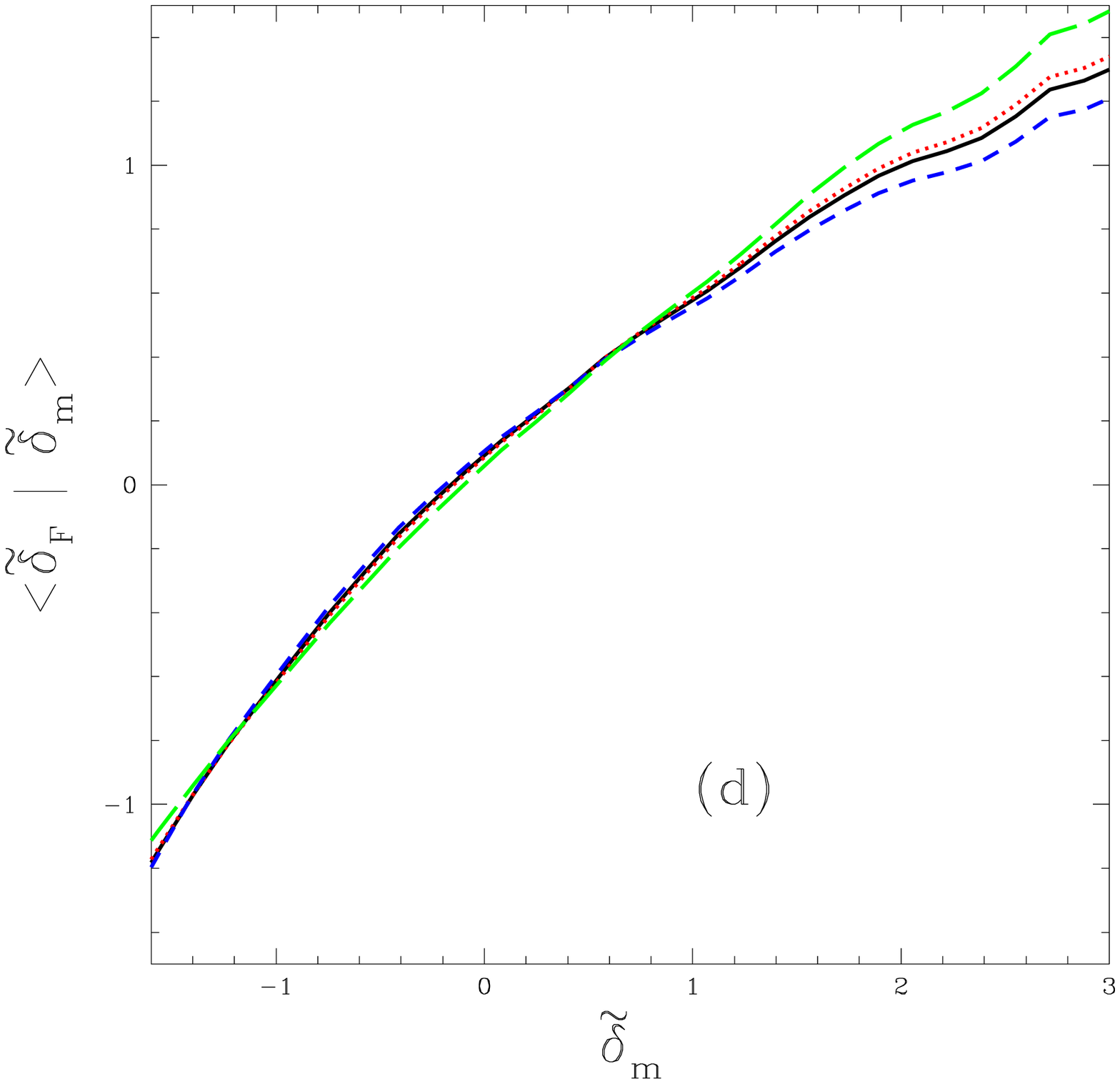}
\end{figure}

  The functions $<\! \dm | \dF \! >$ and $<\! \dF | \dm \! >$ depend
strongly on $\bF$, whereas the functions with the normalized quantities
$<\! \dtm | \dtF \! >$ and $<\! \dtF | \dtm \! >$ have a much weaker
dependence. Essentially, the mean transmitted flux changes the
effective ``bias'' in the \lya forest that determines the value of
$\dF$ that corresponds to a given mass fluctuation, and once a linear
bias is eliminated the dependence on $\bF$ is much smaller.  
Note that Figures \ref{fluxcheck}a,b reflect the fact that the \lya 
forest bias is
larger for higher mean flux decrements (see Figure 10c of McDonald 2002, 
and Croft et al.\ 1999, McDonald et al.\ 2000).  

We give the results for $\bF=0.6$ and $\bF=0.8$ in Table \ref{restab},
as rows 6sd and 6ld, respectively. 
The first row in the Table already contains the results for $\bF=0.67$. 

\subsection{Temperature-density relation}

\begin{figure}
\plotone{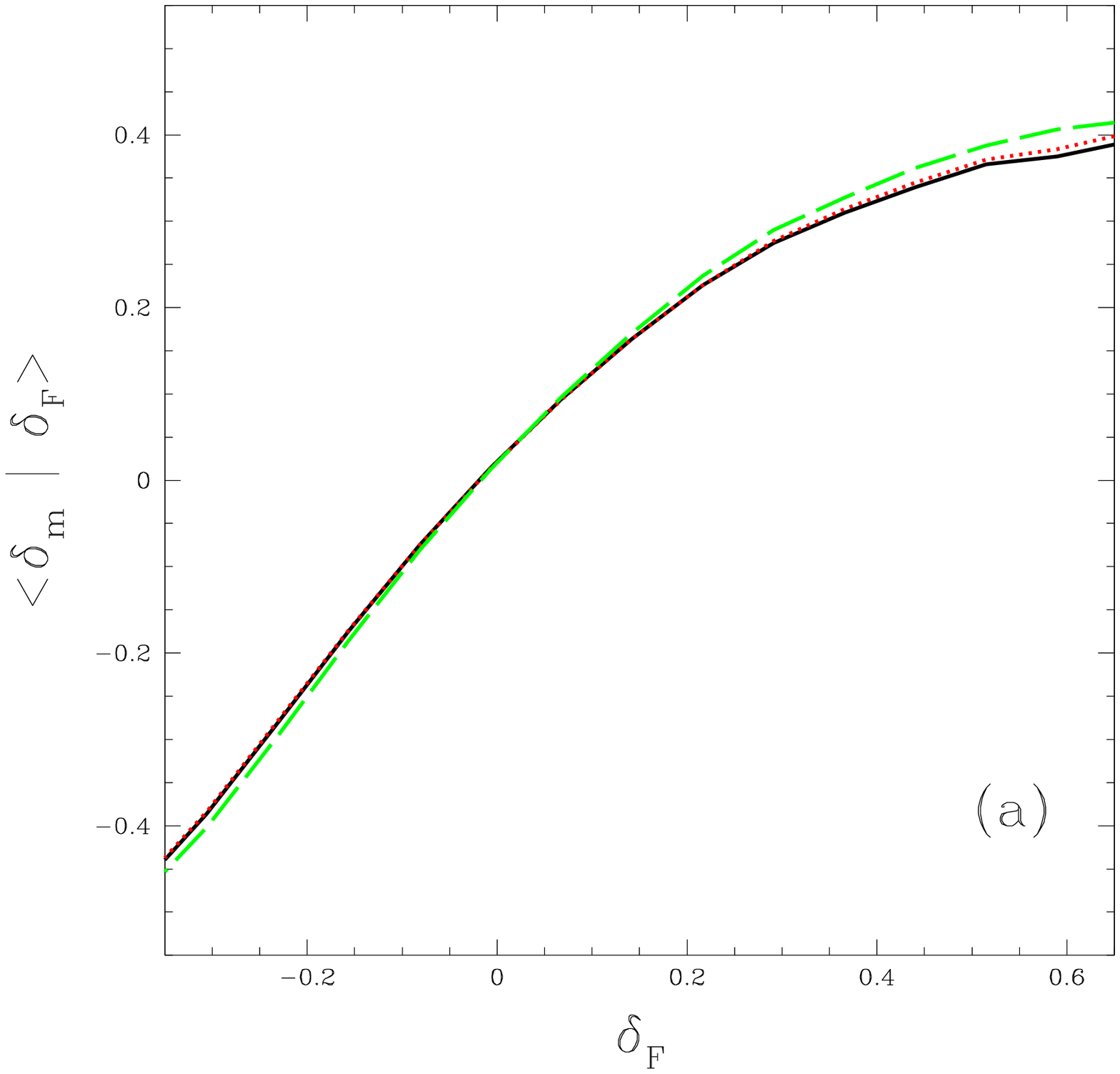}
\caption{Effect of temperature-density relation.  
{\it Black, solid line:} $T_{1.4}=17000$ K, $\gmo=0.3$
(standard values); {\it red, dotted line:}
$T_{1.4}=22000$ K, $\gmo=0.3$; {\it green, dashed line:}
$T_{1.4}=17000$ K, $\gmo=0.6$.}
\label{tdrcheck}
\end{figure}

  The temperature-density relation has practically no effect on the
functions we are analyzing, as shown in Figures \ref{tdrcheck}(a-d).  
\begin{figure}
\plotone{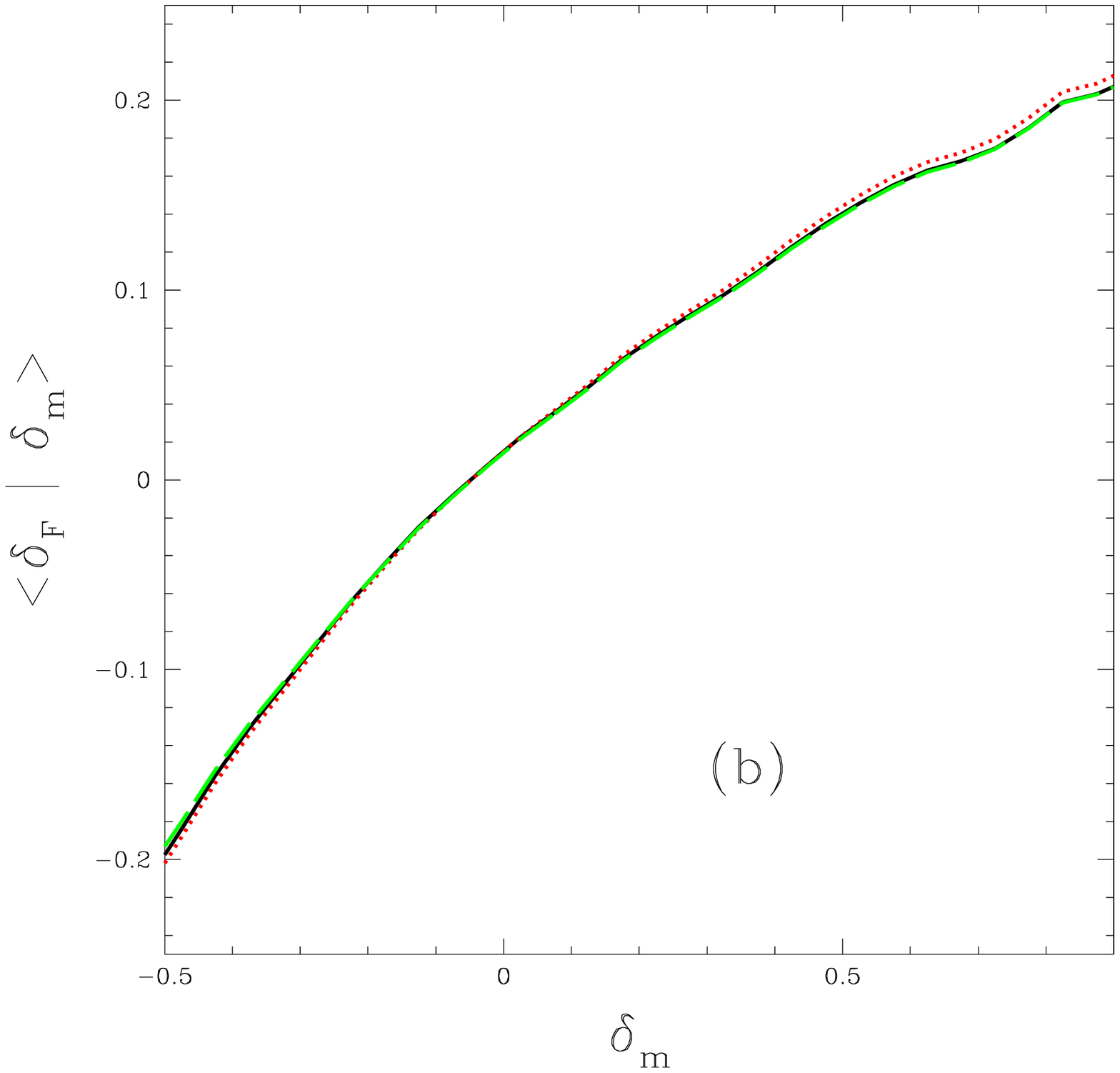}
\end{figure}
\begin{figure}
\plotone{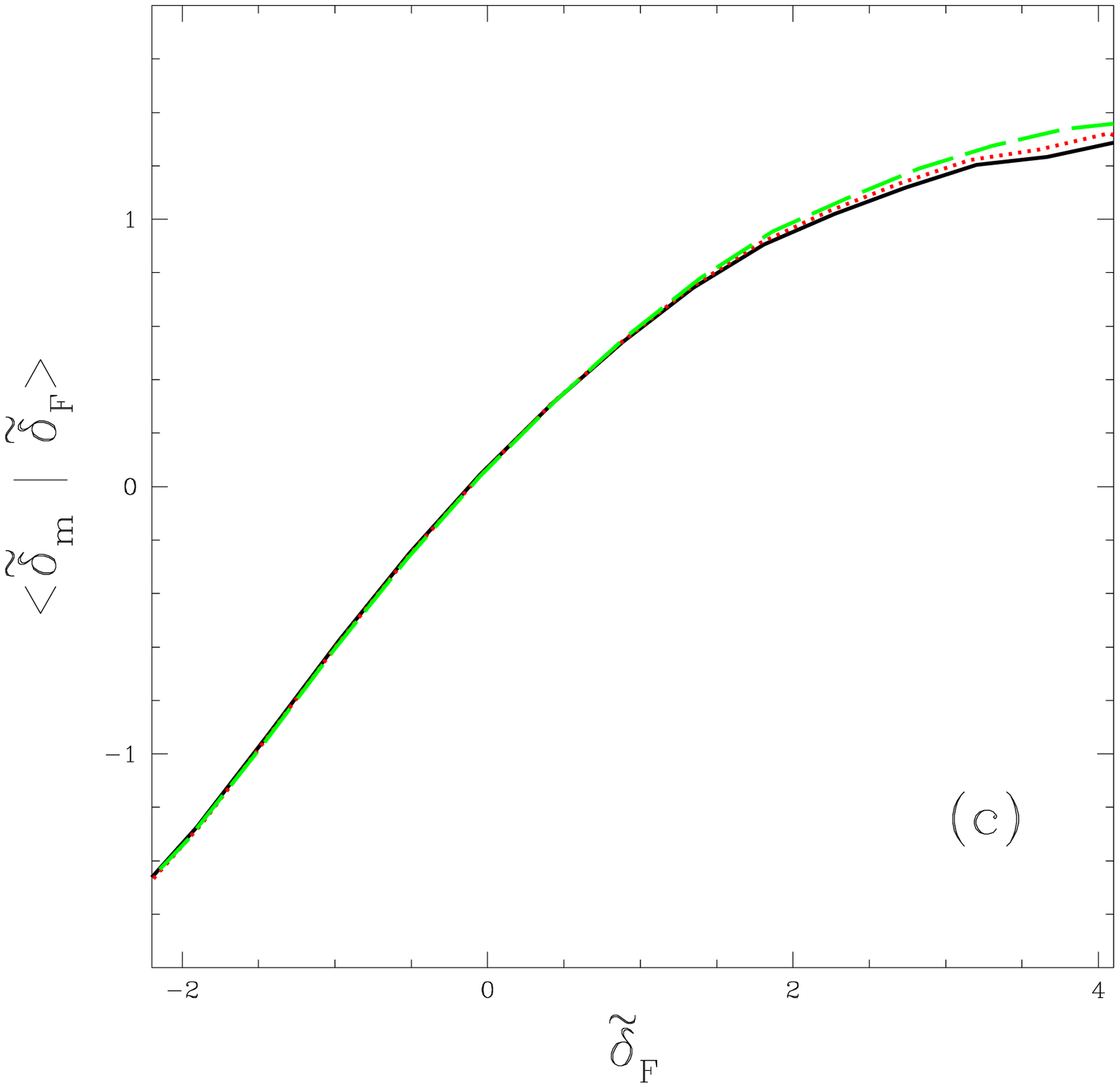}
\end{figure}
\begin{figure}
\plotone{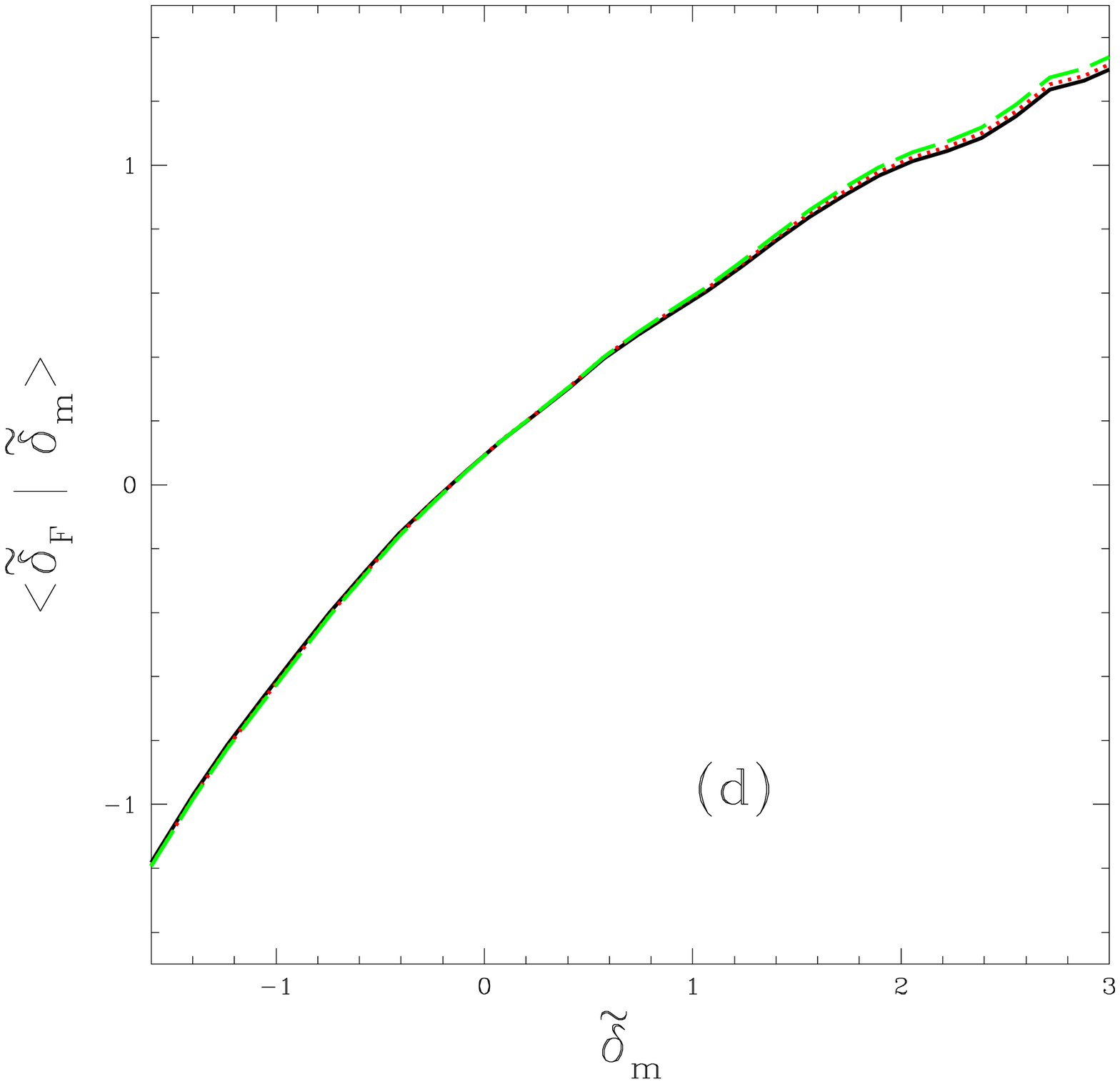}
\end{figure}
The reason for this insensitivity is that the temperature-density
relation does not affect the large-scale properties of the \lya
forest directly, and apparently its effect through the large-scale
bias factor is quite small (see Figure 10b in McDonald 2002).

\subsection{Amplitude of the power spectrum}

\begin{figure}
\plotone{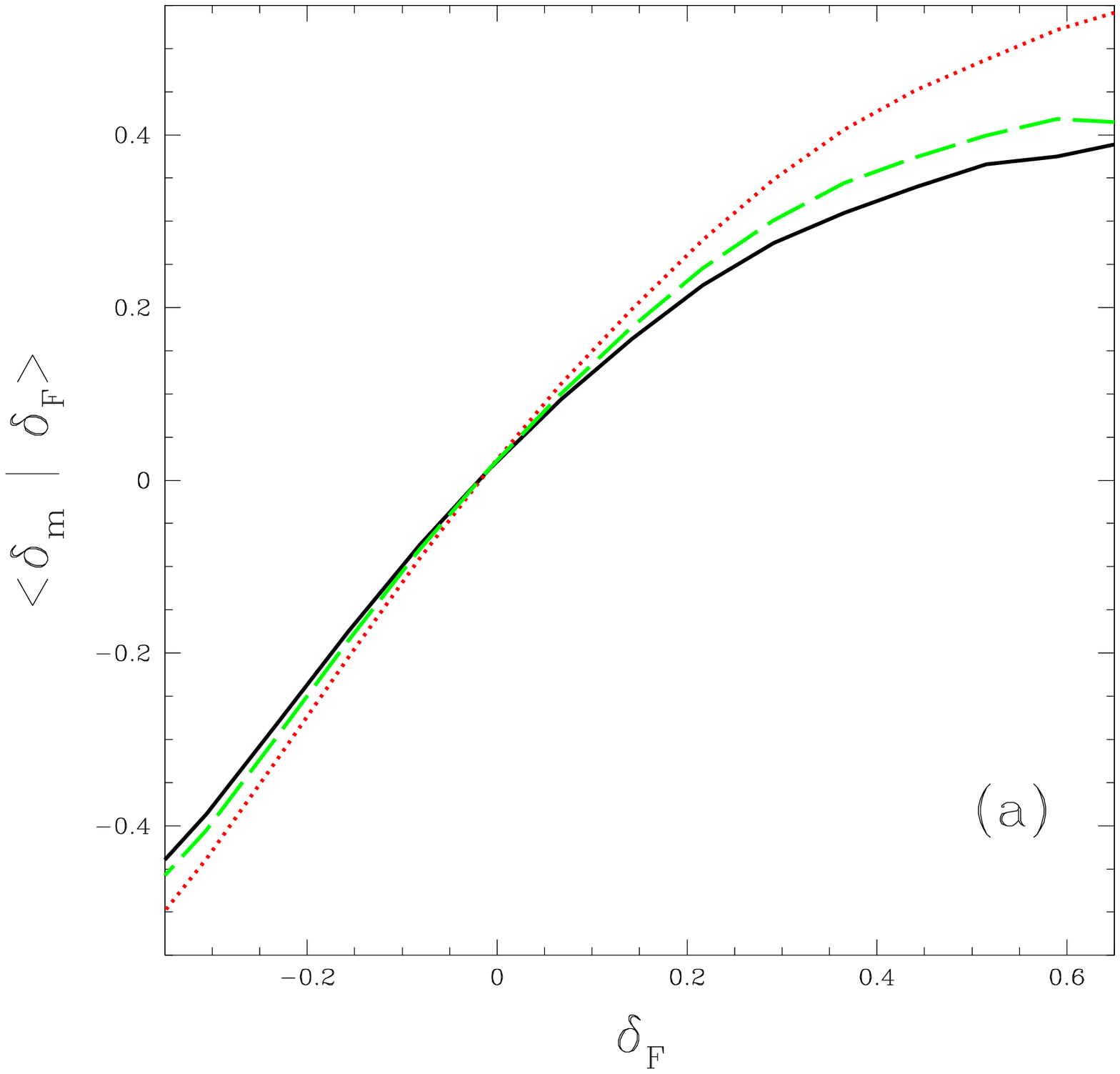}
\caption{Effect of the mass power spectrum.  
{\it Black, solid line:} Standard power spectrum.
{\it Red, dotted line:} rms amplitude increased by 33\%.
{\it Green, dashed line:} $n=0.85$ instead of $n=0.95$,
with fixed amplitude at $k=1~(\hmpc)^{-1}$.}
\label{ampcheck}
\end{figure}

  In Figure \ref{ampcheck},
we show the variations with the amplitude of the mass power
spectrum by increasing the amplitude in our standard model,
shown by the solid line, by 33\%, to produce the dotted line.
In practice, we do this by simply using an output of the same
simulation at a scale factor $1.33$ times larger, but keeping the
mean transmitted flux in the \lya spectrum constant (this is completely
equivalent to running a new simulation with a different amplitude,
except that the effective temperature changes due to the change in
the relation between comoving distance and velocity, but the previous
figures show that the temperature does not affect the results).
\begin{figure}
\plotone{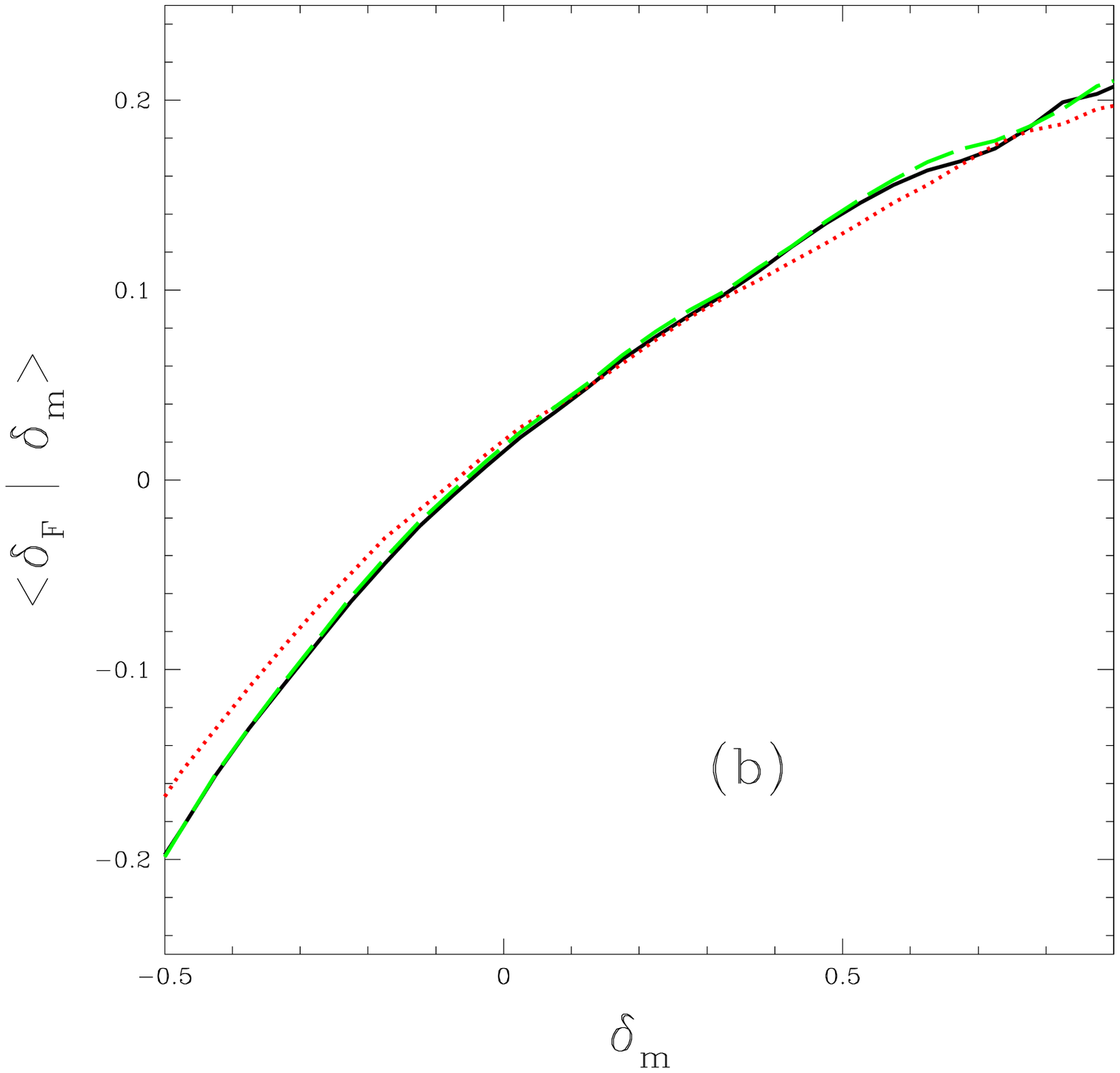}
\end{figure}
\begin{figure}
\plotone{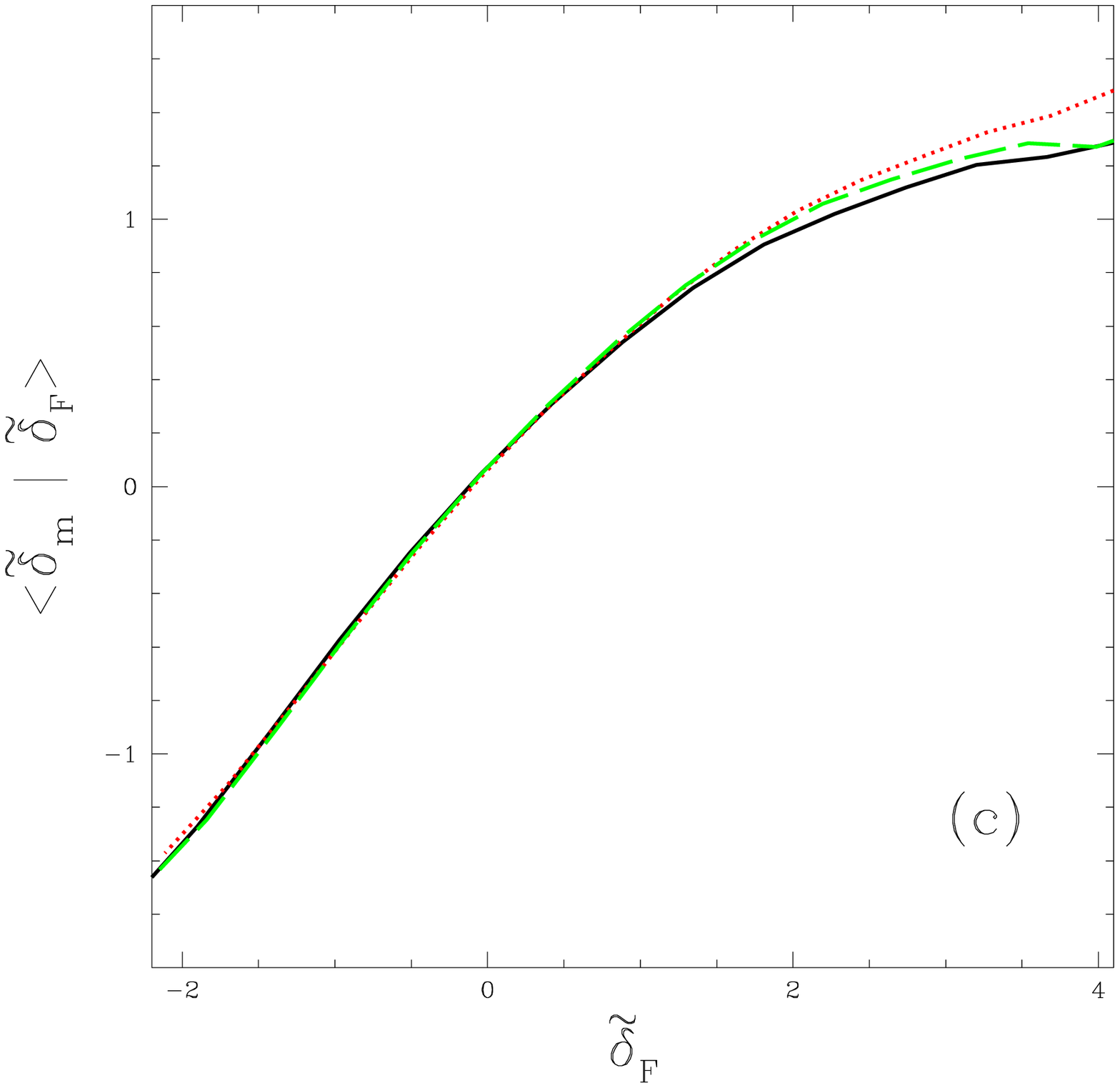}
\end{figure}
\begin{figure}
\plotone{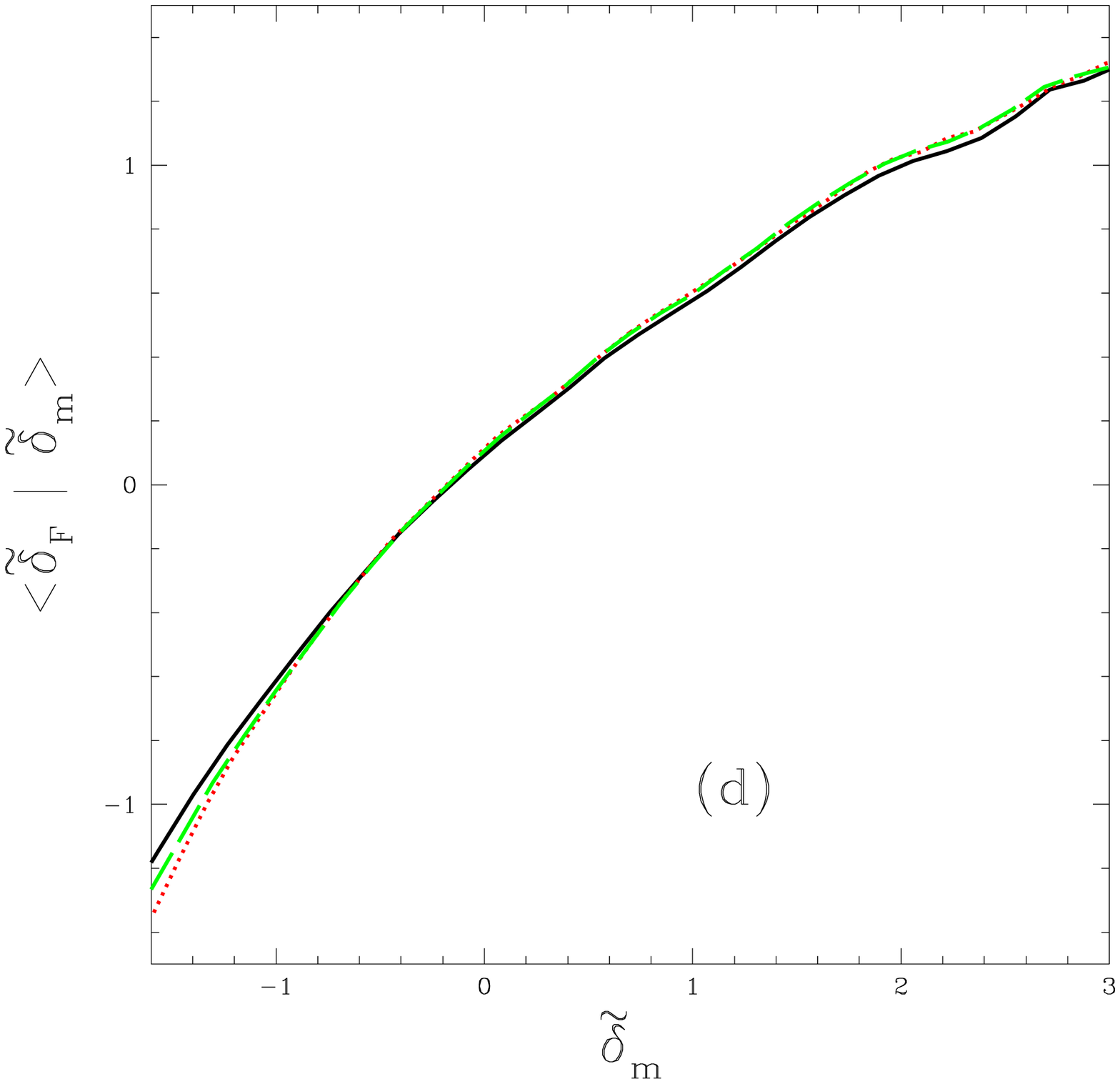}
\end{figure}
The result reflects that the bias factor of the \lya forest decreases
with increasing power spectrum amplitude (see Figure 10a of McDonald
2002). The difference is partially cancelled when using the
normalized variables $\dtF$ and $\dtm$, although less so for the
function $<\dtF | \dtm>$.

 We also vary the power-law slope of the power spectrum in Figure
\ref{ampcheck}, by changing
$n$ from 0.95 to 0.85 while holding the power fixed at wavenumber
$k=1 \hmpc$.  The effect we see is relatively small. Had we carefully
chosen the value of $k$ at which we hold the power fixed to minimize
the variation with $n$, the effect might be even smaller.

We give the results for $\sigma_8=1$ and $n=0.85$ in Table \ref{restab},
as rows 8d and 8ld, respectively. 
The first row in the Table is for the standard power spectrum. 

\section{DISCUSSION}

  We have presented predictions for the expected correlation of the
mass and the \lya forest transmitted flux, smoothed over a cube size
of $\sim 10 \hmpc$. Our most basic result is that the relation between
$<\! \dtm | \dtF\! >$ and $\dtF$ is linear over most of the range of
$\dtF$ (excluding rare, high $\dtF$ values), with a slope of 0.6 on
this smoothing scale.
We have shown that this correlation is not sensitive
to the temperature-density relation of the gas. There is a dependence
on the mean transmitted flux and the amplitude of the mass power
spectrum, but these quantities are already measured to reasonable
accuracy (Croft \etal 1999, McDonald \etal 2000). We have also shown
that the predictions do not suffer from uncertainties due to the
resolution of the simulations. 
Calculations with even larger boxes than
used here will be desirable, because the uncertainties in the
predictions due to the variance in the simulations and the suppression
of the large-scale power may still be significant; 
however, this uncertainty
is mostly isolated in the most rare, high density regions.

  A similar relation holds between $<\! \dtF | \dtm\! >$ and $\dtm$.
We notice, however, that the observational determination of this other
function will be affected by galaxy shot noise. The number of galaxies
observed in a certain cube in redshift space containing a fixed mass is
subject to shot noise, and this inevitably introduces a smoothing of the
function $<\! \dtF | \dtm\! >$, which must be taken into account before
any comparisons to our theoretical results are made (in contrast, galaxy
shot noise does not change the average $<\! \dF | \dm \! >$, but it does
alter the rms dispersion $\sigma_F$ needed to compute $\dtF$). Measuring
the effects of galaxy shot noise can also teach us useful information
about how galaxies form, because galaxy shot noise does not generally
need to be strictly described by Poisson statistics. For example, for a
fixed mass contained within a cube, once a galaxy is found in the cube
the probability to find others may be lower because some mass has
already been used up by that galaxy.

  Comparing these theoretical predictions with observations allows
for several tests of the basic theory of the \lya forest, and can
reveal new information on the spatial distribution of the galaxies.
The functions $<\! \dtm | \dtF \! >$ and $<\! \dtF | \dtm \! >$, which
should not be affected by any linear galaxy bias, provide a powerful
test of the assumption that the basic framework assumed here for the
nature of the \lya forest is correct, and that galaxies trace the mass
on large scales apart from linear bias. If the slope of
$<\! \dtm | \dtF \! >$ as a function of $\dtF$ were found to be larger
than predicted, this would imply that the \lya forest is much more
closely associated with galaxies than is expected from their common
correlation with the mass distribution.
If the observed slope were smaller than predicted, it would indicate
that the \lya absorbing gas and the galaxies tend to avoid each other
for some reason.

  If the correlation of $\dtm$ and $\dtF$ is as expected, this will
imply a strong confirmation of the basic model we have for the \lya
forest, and will justify using the \lya forest as a predictor of the
mass fluctuations. Comparing the predicted function 
$<\! \dm | \dF \! >$ with the observed $<\! \dg | \dF \! >$ will then
yield the bias factor of any type of galaxies for which these
observations can be made.

  The correlation of galaxies and the \lya forest can be measured as a
function of scale. Our predicted value of 0.6 for the slope of
$< \! \dtm | \dtF \! > $ as a function of $\dtF$ should decrease with
the smoothing scale, in a way that reflects the shape of the mass
autocorrelation function. This will allow a precise test of the idea
that the large-scale distributions of different types of galaxies differ
only in a constant linear bias relative to the mass fluctuations.

\acknowledgements
  We acknowledge useful discussions with Kurt Adelberger, Charles
Steidel, and David Weinberg.  We thank Nick Gnedin for the HPM 
code.
PM is supported by a grant from the Packard Foundation.
RC is supported by grants AST93-18185 and ASC97-40300.
JM is supported by grant NSF-0098515.

\begin{deluxetable}{lcccccccccc}
\tablecolumns{11}
\tablecaption{Selected Results\label{restab}}
\tablehead{ \colhead{Fig.,}&&&\colhead{$<\dtm | \dtF>$}&&&&
\colhead{$<\dtF | \dtm>$}&&&\\
\colhead{line} & \colhead{$\sigma_F$} & \colhead{$\sigma_m$} &
\colhead{$\dtF=-1.5$} & \colhead{0} & 
\colhead{1.5} & \colhead{3} & 
\colhead{$\dtm=-1$} & \colhead{0} & 
\colhead{1} & \colhead{2}}
\startdata
2, s \tablenotemark{a} & 0.16 & 0.30 & -0.97 & 0.07 & 0.80 & 1.17 & -0.61 & 0.09 & 0.58 & 1.00\\
2, ld \tablenotemark{b} & 0.15 & 0.26 & -0.95 & 0.09 & 0.73 & 1.03 & -0.56 & 0.09 & 0.53 & 0.94\\
2, sd \tablenotemark{c} & 0.18 & 0.36 & -1.00 & 0.05 & 0.87 & 1.35 & -0.68 & 0.12 & 0.63 & 1.05\\
6, ld \tablenotemark{d} & 0.11 & 0.30 & -1.09 & 0.10 & 0.86 & 1.26 & -0.63 & 0.06 & 0.60 & 1.11\\
6, sd \tablenotemark{e} & 0.19 & 0.30 & -0.92 & 0.06 & 0.77 & 1.12 & -0.60 & 0.11 & 0.56 & 0.94\\
8, d \tablenotemark{f} & 0.18 & 0.39 & -0.98 & 0.06 & 0.84 & 1.27 & -0.65 & 0.12 & 0.60 & 1.03\\
8, ld \tablenotemark{g} & 0.17 & 0.33 & -1.00 & 0.07 & 0.84 & 1.21 & -0.64 & 0.11 & 0.60 & 1.03\\
\enddata
\tablecomments{First column gives the Figure and curve to which the 
row of results applies.  Solid line=s,
dotted=d, long-dashed=ld, and short-dashed=sd.  Value
for $<\dtm | \dtF=3>$ underestimated due to limited box size.  
$\sigma_m$ and $\sigma_F$ intended only for use as conversion
factors from $\tilde{\delta}$ to $\delta$ -- alone they may be 
sensitive to resolution and box size.}
\tablenotetext{a}{Standard case.}
\tablenotetext{b}{$12\hmpc$ cube, $4.2\hmpc$ Gaussian.}
\tablenotetext{c}{$8\hmpc$ cube, $2.7\hmpc$ Gaussian.}
\tablenotetext{d}{$\bar{F}=0.8$.}
\tablenotetext{e}{$\bar{F}=0.6$.}
\tablenotetext{f}{$\sigma_8=1$}
\tablenotetext{g}{$n=0.85$.}

\end{deluxetable}


\newpage

\begin{references}

\reference{} Adelberger, et al. 2001, in preparation

\reference{} Bechtold, J., Crotts, A. P. S.,
               Duncan, R. C., \& Fang, Y. 1994, ApJ, 437, L83

\reference{} Bergeron, J., \& Boiss\'e, P. 1991, A\& A, 243, 344

\reference{} Bi, H. G. 1993, ApJ, 405, 479

\reference{} Bi, H. G. \& Davidsen, A. F. 1997, ApJ, 479, 523

\reference{} Cen, R. Miralda-Escud\'e, J., Ostriker, J. P., \& 
             Rauch, M. 1994, ApJ, 437, L9
             
\reference{} Cen, R., Ostriker, J.P., Prochaska, J.X., Wolfe, A.M. 2001, 
             in preparation             

\reference{} Chen, H.-W., Lanzetta, K. M., Webb, J. K., \& Barcons, X. 2001,
             ApJ, 559, 654    
              
\reference{} Croft, R. A. C., Weinberg, D. H., Bolte, M., Burles, S., 
             Hernquist, L., Katz, N., Kirkman, D., \& Tytler, D. 
             2002, ApJ, submitted (astro-ph/0012324)
             
\reference{} Croft, R. A. C., Weinberg, D. H., Pettini, M., 
             Hernquist, L., \& Katz, N. 1999, ApJ, 520, 1

\reference{} Dinshaw, N., Impey, C. D., Foltz, C. B.,
             Weymann, R. J., \& Chaffee, F. H. 1994, ApJ, 437, L87
          
\reference{} Dinshaw, N., Weymann, R. J., Impey, C. D., Foltz, C. B., 
             Morris, S. L., \& Ake, T. 1997 ApJ, 491, 45
             
\reference{} Dolan, J. F., Michalitsianos, A. G., Nguyen, Q. T., \&
             Hill, R. J. 2000, ApJ, 539, 111             

\reference{} Gnedin, N. Y. \& Hui, L. 1998, MNRAS, 296, 44 

\reference{} Hernquist, L., Katz, N.,
             Weinberg, D. H., \& Miralda-Escud\'e, J. 1996, ApJ, 457, L51 	
  
\reference{} Lanzetta, K. M., Bowen, D. V., Tytler, D., \& Webb, J. K.
             1995, ApJ, 442, 538
             
\reference{} L\'opez, S., Hagen, H.-J., \& Reimers, D. 
             2000, A\&A, 357, 37             
             
\reference{} McDonald, P. 2002, ApJ, submitted (astro-ph/0108064)

\reference{} McDonald, P., Miralda-Escud\'e, J., Rauch, M.,
             Sargent, W. L. W., Barlow, T. A., Cen, R., \&
             Ostriker, J. P. 2000, ApJ, 543, 1

\reference{} McDonald, P., Miralda-Escud\'e,
         J., Rauch, M., Sargent, W. L. W., Barlow, T. A.,\& Cen, R.
         2002, ApJ, in press (astro-ph/0005553)

\reference{} Miralda-Escud\'e, J., 
         Cen, R., Ostriker, J. P., \& Rauch, M. 1996, ApJ, 471, 582

\reference{} Monier, E. M., Turnshek, D. A., \& Hazard, C. 1999, ApJ,
             522, 627

\reference{} Penton, S. V., Stocke J. T., \& Shull, J. M. 2001,
             ApJ, in press (astro-ph/0109277) 

\reference{} Petitjean, P., Surdej, J., Smette, A., Shaver, P., 
             Muecket, J., \& Remy, M. 1998, A\&A, 334, L45
             
\reference{} Primack, J. R. 2000, preprint (astro-ph/0007187)	 

\reference{} Rauch, M., Miralda-Escud\'e, J., 
             Sargent, W. L. W., 
             Barlow, T. A., Weinberg, D. H., Hernquist, L., Katz, 
             N., Cen, R., \& Ostriker, J. P. 1997, ApJ, 489, 7

\reference{} Steidel, C. C., Dickinson, M., \& Persson, E. 1994, 
               ApJ, 437, L75

\reference{} Zhang, Y., Anninos, P., \& Norman, M. L. 1995, ApJ, 453, L57	 

\reference{} Zhang, Y., Meiksin, A., Anninos, P., \& Norman, M., L.
             1998, ApJ, 495, 63

\end{references}
\end{document}